%% file: main.tex
\documentclass[acmsmall,nonacm]{acmart}

\usepackage{booktabs} 
\usepackage[dvipsnames]{xcolor}
\usepackage{subcaption}

\usepackage{listings}
\usepackage{placeins} 

\definecolor{codegreen}{rgb}{0,0.6,0}
\definecolor{codegray}{rgb}{0.5,0.5,0.5}
\definecolor{codepurple}{rgb}{0.58,0,0.82}
\definecolor{backcolour}{rgb}{0.95,0.95,0.92}

\lstdefinestyle{mystyle}{
    backgroundcolor=\color{backcolour},   
    commentstyle=\color{codegreen},
    keywordstyle=\color{magenta},
    numberstyle=\tiny\color{codegray},
    stringstyle=\color{codepurple},
    basicstyle=\ttfamily\footnotesize,
    breakatwhitespace=false,         
    breaklines=true,                 
    captionpos=b,                    
    keepspaces=true,                 
    numbers=left,                    
    numbersep=5pt,                  
    showspaces=false,                
    showstringspaces=false,
    showtabs=false,                  
    tabsize=2
}

\lstdefinestyle{myHIPStyle}{
	frame=tb,
	language=[ANSI]C++,
	showstringspaces=false,
	basicstyle=\footnotesize\ttfamily,
	identifierstyle=\color{Maroon},
	keywordstyle=\color{MidnightBlue},
	numberstyle=\color{Emerald},
	stringstyle=\color{OliveGreen},
	commentstyle=\itshape\color{gray},
	emph={
	    hipMalloc,
	    cudaMalloc,
	    hipFree, 
	    cudaFree, 	    
	    hipMemcpy,
	    cudaMemcpy,
	    hipMemcpyHostToDevice,
	    cudaMemcpyHostToDevice,	    
	    hipMemcpyDeviceToHost,
	    cudaMemcpyDeviceToHost,	    
	    hipMemcpyAsync,
	    cudaMemcpyAsync,	    
	    hipDeviceSynchronize,
	    cudaDeviceSynchronize,	    
	    hipLaunchKernelGGL,
	    hipStream_t,
	    cudaStream_t,	    
	    hipStreamCreate,
	    cudaStreamCreate,	    
	    hipStreamDestroy,
	    cudaStreamDestroy,	    
	    hipStreamSynchronize, 
	    cudaStreamSynchronize, 	    
		hipGetDeviceCount,
		cudaGetDeviceCount,		
		hipDeviceProp_t,
		cudaDeviceProp,		
		hipGetDeviceProperties,
		cudaGetDeviceProperties,
		hipGetErrorString,		
		cudaGetErrorString,
		hipLaunchKernel,		
		cudaLaunchKernel,
		hipSuccess,
		cudaSuccess,
		cuda_runtime.h,
	    __global__, 
	    __shared__,
	    uint32_t,
	},
	emphstyle={\color{MidnightBlue}},
	breaklines=true,
	tabsize=2,
	numbers=left,
	numbersep=6pt,
	captionpos=t,
}

\lstdefinestyle{myHIPStyle2}{
    frame=tb,
    language=[ANSI]C++,
    backgroundcolor=\color{white},   
    commentstyle=\color{codegreen},
    basicstyle=\ttfamily\footnotesize,
    keywordstyle=\color{MidnightBlue},
    stringstyle=\color{OliveGreen},
    numberstyle=\color{Emerald},
    commentstyle=\itshape\color{Green},
    morekeywords={
        __device__,
        __global__, 
	    __shared__,
        constexpr,
        size_t,
        uint8_t,
        uint32_t,
        uint3,
        float3, 
        float4,
        threadIdx,
        blockDim,
        blockIdx,
    },
    deletekeywords={f,F},
	emph={
        hipInit,
        hipDevice,
        hipDeviceGet,
        hipCtx,
        hipCtxCreate, 
        oroCtx,
        oroCtxCreate,
        oroGetDeviceCount,
        oroInitialize,
        oroInit,
        oroDevice,
        oroDeviceGet,
        oroDeviceProp,
        oroGetDeviceProperties,
        oroMalloc,
        oroMemcpyHtoD,
        oroDeviceptr,
        dot,
        cross,
        lerp,
        normalize,
        sqrtf,
        cosf,
        sinf,
        fmax,
        oroGetRawCtx,
        oroGetRawDevice,
        hiprtContextCreationInput,
        hiprtContext,
        hiprtSetLogLevel,
        hiprtCreateContext, 
        hiprtTriangleMeshPrimitive,
        hiprtGeometryBuildInput,
        hiprtBuildOptions,
        hiprtContextCreationInput,
        hiprtCreateContext,
        hiprtSetLogLevel,
        hiprtTriangleMeshPrimitive,
        hiprtGeometryBuildInput,
        hiprtBuildOptions,
        hiprtGetGeometryBuildTemporaryBufferSize,
        hiprtDevicePtr,
        hiprtCreateGeometry,
        hiprtBuildGeometry,
        hiprtAABBListPrimitive,
        hiprtSceneBuildInput,
        hiprtInstance,
        hiprtFrameSRT,
        hiprtGetSceneBuildTemporaryBufferSize,
        hiprtCreateScene,
        hiprtBuildScene,
        hiprtFuncDataSet,
        hiprtFuncTable,
        hiprtCreateFuncTable,
        hiprtSetFuncTable,
        hiprtRay,
        hiprtSceneTraversalAnyHitCustomStack,
        hiprtSceneTraversalClosestCustomStack,
        hiprtHit,
        hiprtSharedStackBuffer,
        hiprtGlobalStackBuffer,
        hiprtGlobalStack,
        hiprtEmptyInstanceStack,
        hiprtBuildTraceKernels,
        hiprtFuncNameSet,
        hiprtApiFunction,
        hiprtGeometry,
        hiprtScene,
	},
	emphstyle=\color{Maroon},
    breakatwhitespace=false,         
    breaklines=true,                 
    keepspaces=true,                 
    numbers=left,       
    numbersep=5pt,                  
    showspaces=false,                
    showstringspaces=false,
    showtabs=false,                  
    tabsize=2,
}

\lstdefinestyle{myBASHStyle}{
	frame=tb,
	language=Bash,
	showstringspaces=false,
	basicstyle=\footnotesize\ttfamily,
	identifierstyle=\color{Black},
	keywordstyle=\color{Black},
	numberstyle=\color{Black},
	stringstyle=\color{Black},
	commentstyle=\itshape\color{gray},
	emph={
	},
	emphstyle={\color{MidnightBlue}},
	breaklines=true,
	tabsize=2,
	numbers=left,
	numbersep=6pt,
	captionpos=t,
}

\lstdefinestyle{myCDNAStyle}{
	frame=tb,
	language=[x86masm]Assembler,
	showstringspaces=false,
	basicstyle=\lstfont{black},
	identifierstyle=\lstfont{Maroon},
	keywordstyle=\lstfontbold{MidnightBlue},
	numberstyle=\lstfont{Emerald},
	stringstyle=\lstfont{OliveGreen},
	commentstyle=\lstfont{gray},
	emph={
	},
	emphstyle={\lstfontbold{MidnightBlue}},
	breaklines=true,
	tabsize=2,
	numbers=left,
	captionpos=t,
}

\usepackage[normalem]{ulem}

\newcommand{\articleitself}{technical report}

\newcommand\blfootnote[1]{%
  \begingroup
  \renewcommand\thefootnote{}\footnote{#1}%
  \addtocounter{footnote}{-1}%
  \endgroup
}

\copyrightyear{2021}
\acmYear{2021}
\setcopyright{none}
\acmConference[Advanced Micro Devices, Inc. Technical Report]{Advanced Micro Devices, Inc. Technical Report}{December 14}

\setcitestyle{square}

\begin{document}


\title{Ray Tracing using HIP} 
\author{Atsushi Yoshimura}
\affiliation{%
  \institution{\normalsize{Advanced Micro Devices, Inc.}}\country{Japan}}

\author{Kenta Eto}
\affiliation{%
  \institution{\normalsize{Advanced Micro Devices, Inc.}}\country{Japan}}

\author{Daniel Meister}
\affiliation{%
  \institution{\normalsize{Advanced Micro Devices, Inc.}}\country{Japan}}

\author{Takahiro Harada}
\affiliation{%
  \institution{\normalsize{Advanced Micro Devices, Inc.}}\country{USA}}




\renewcommand{\shortauthors}{Yoshimura et al.}

\blfootnote{email: \{ Atsushi.Yoshimura, Kenta.Eto, Daniel.Meister, Takahiro.Harada \}@amd.com \\ Advanced Micro Devices, Inc. Technical Report No. 26-01-4793. v.1, January 10, 2026. }

\maketitle





\input{sec/1_intro}
\input{sec/2_orochi}
\input{sec/3_rendering_triangles}
\input{sec/4_ambient_occlusion}
\input{sec/5_path_tracing}
\input{sec/6_bvh}
\input{sec/7_hiprt}
\input{sec/8_next_steps}

\bibliographystyle{plainnat}
\bibliography{main}

\end{document}

%% file: sec/1_intro.tex
\section{Introduction}
\begin{figure}
    \begin{tabular}[b]{@{}c@{}}
    \includegraphics[width=.45\linewidth]{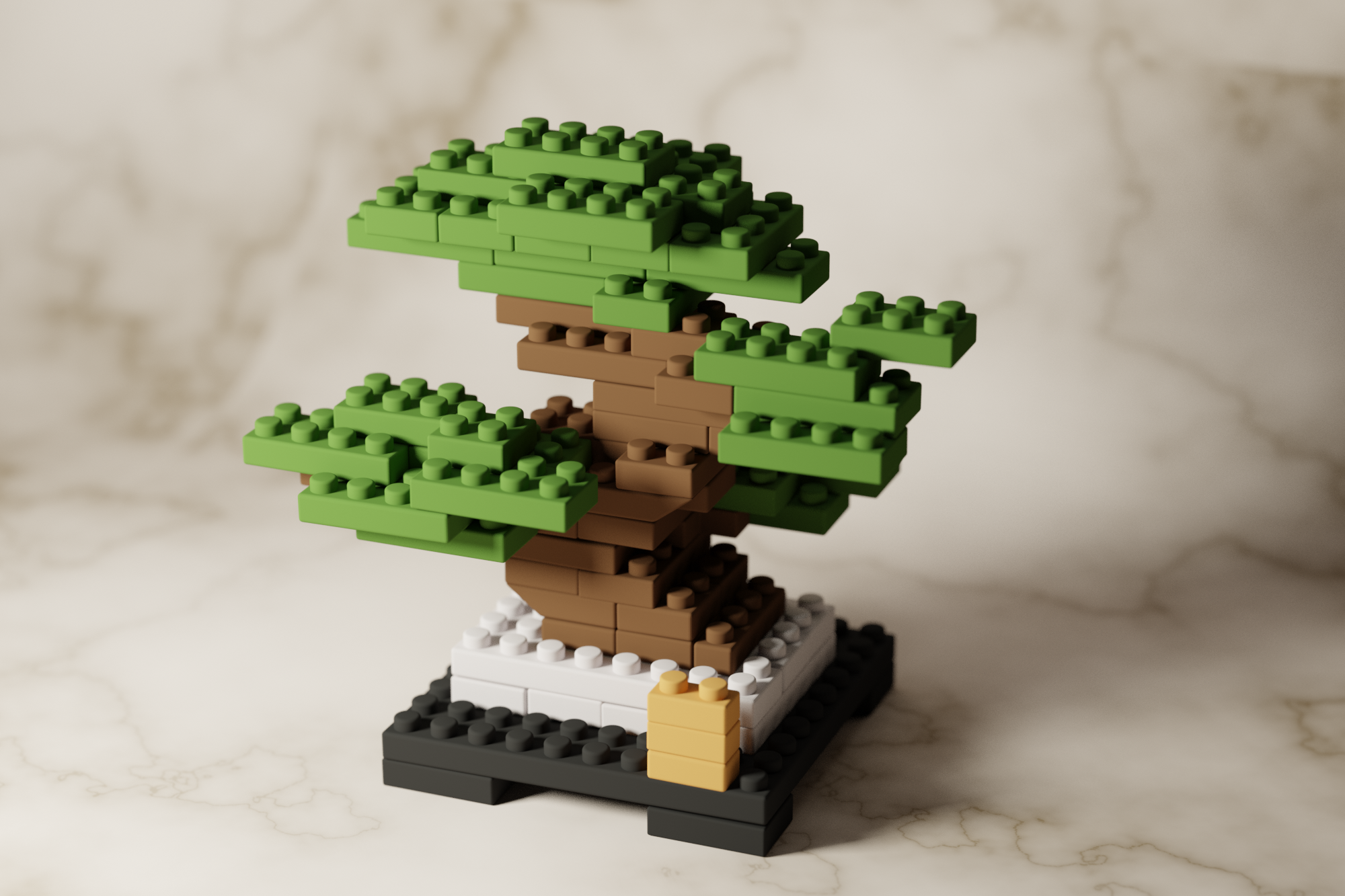}
    \includegraphics[width=.45\linewidth]{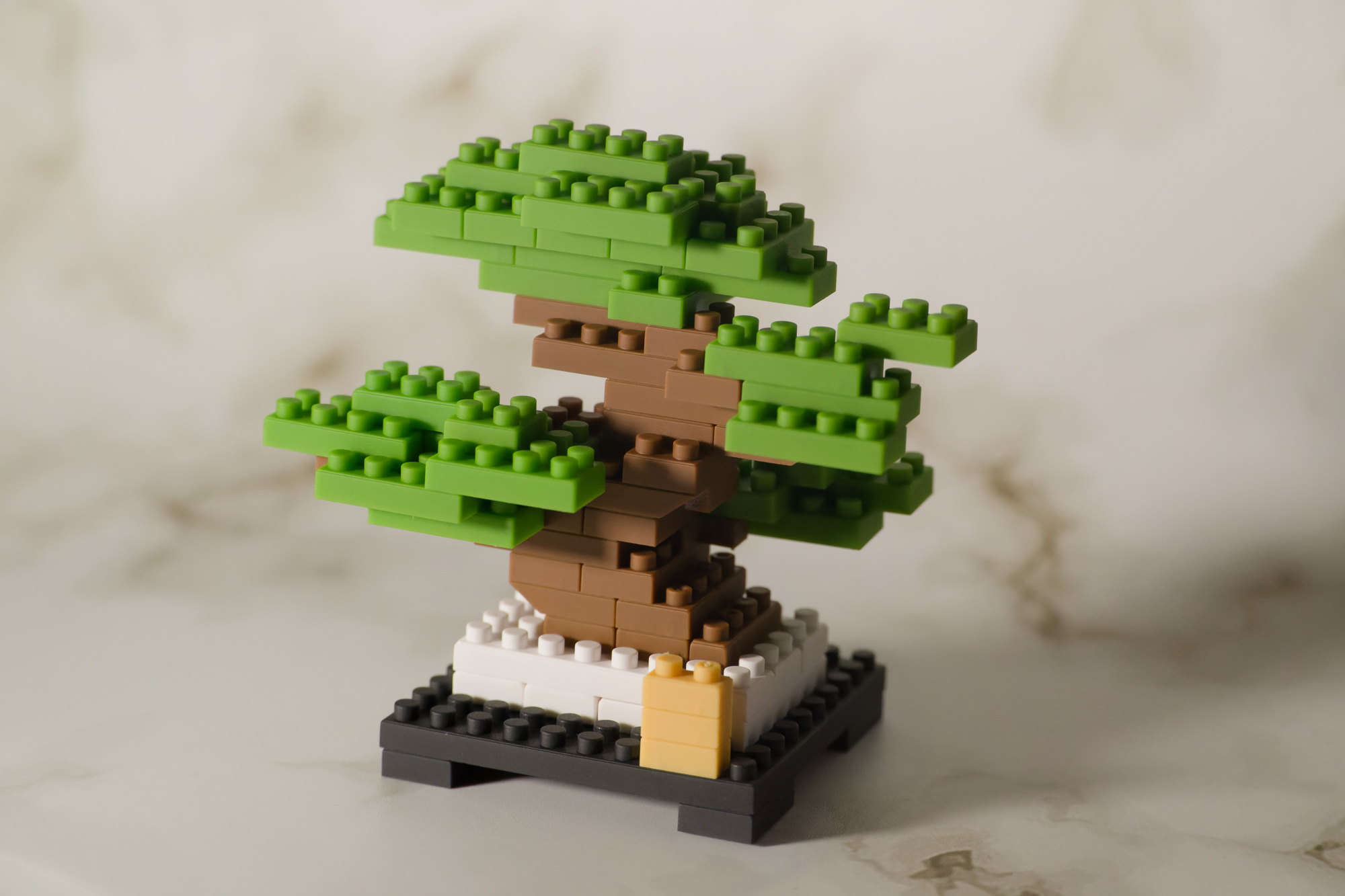}
    \end{tabular}%
    
    \caption{Ray tracing or photo? }
    \label{fig:RT_or_photo}
  \end{figure}

\begin{figure}[t]
    \begin{tabular}[b]{@{}c@{}}
    \includegraphics[width=.225\linewidth]{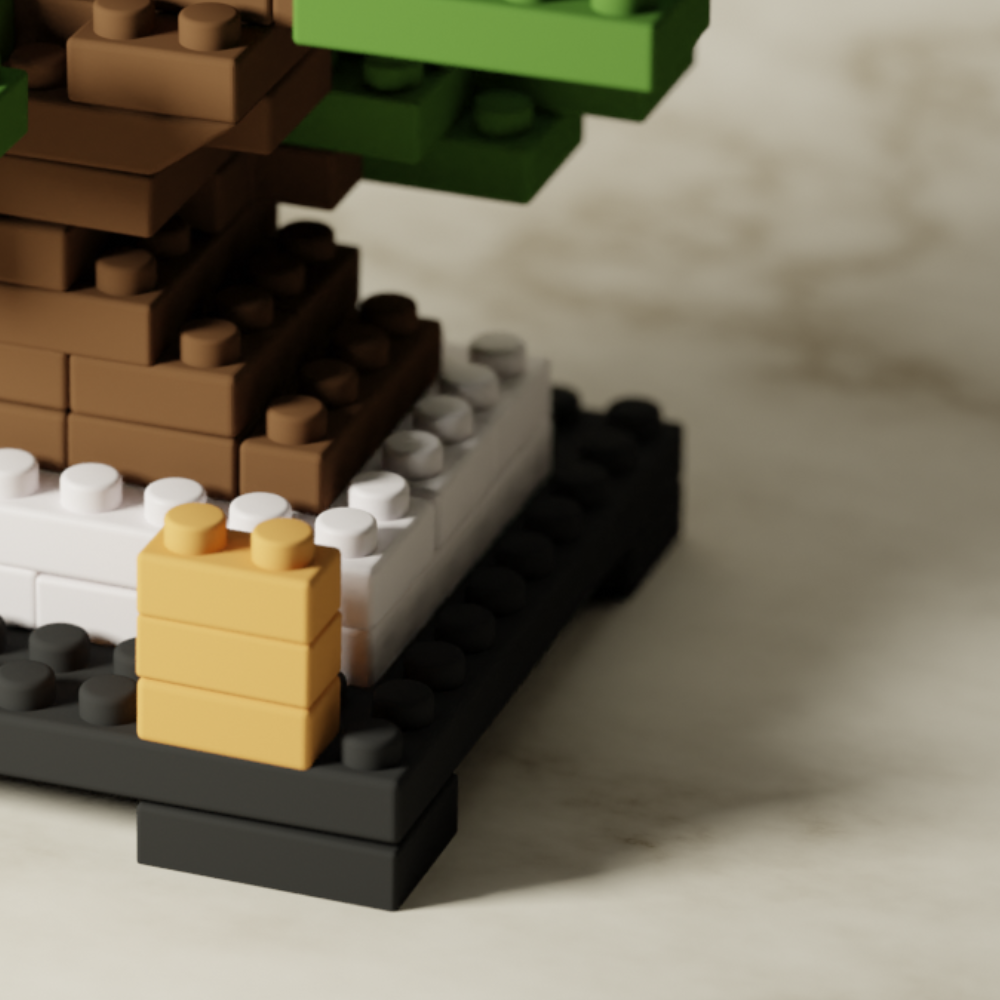}
    \includegraphics[width=.225\linewidth]{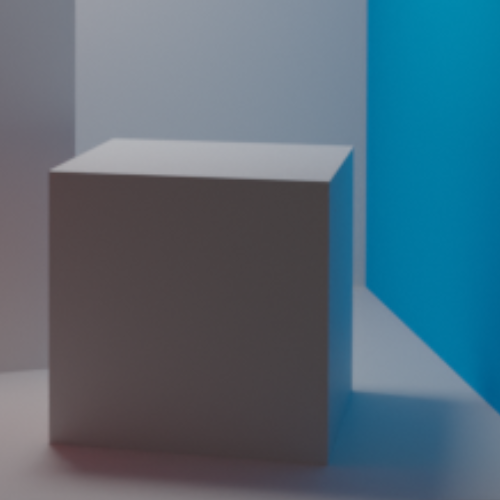}
    \includegraphics[width=.225\linewidth]{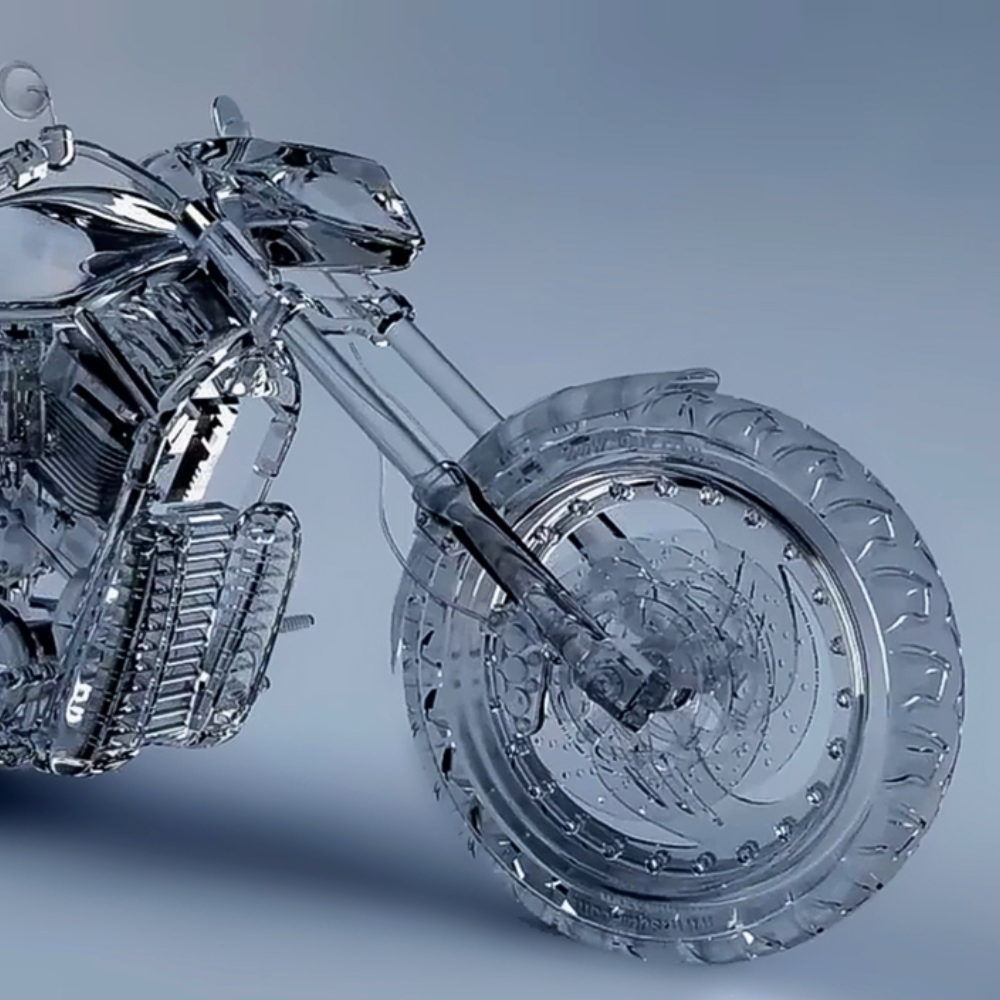}
    \includegraphics[width=.225\linewidth]{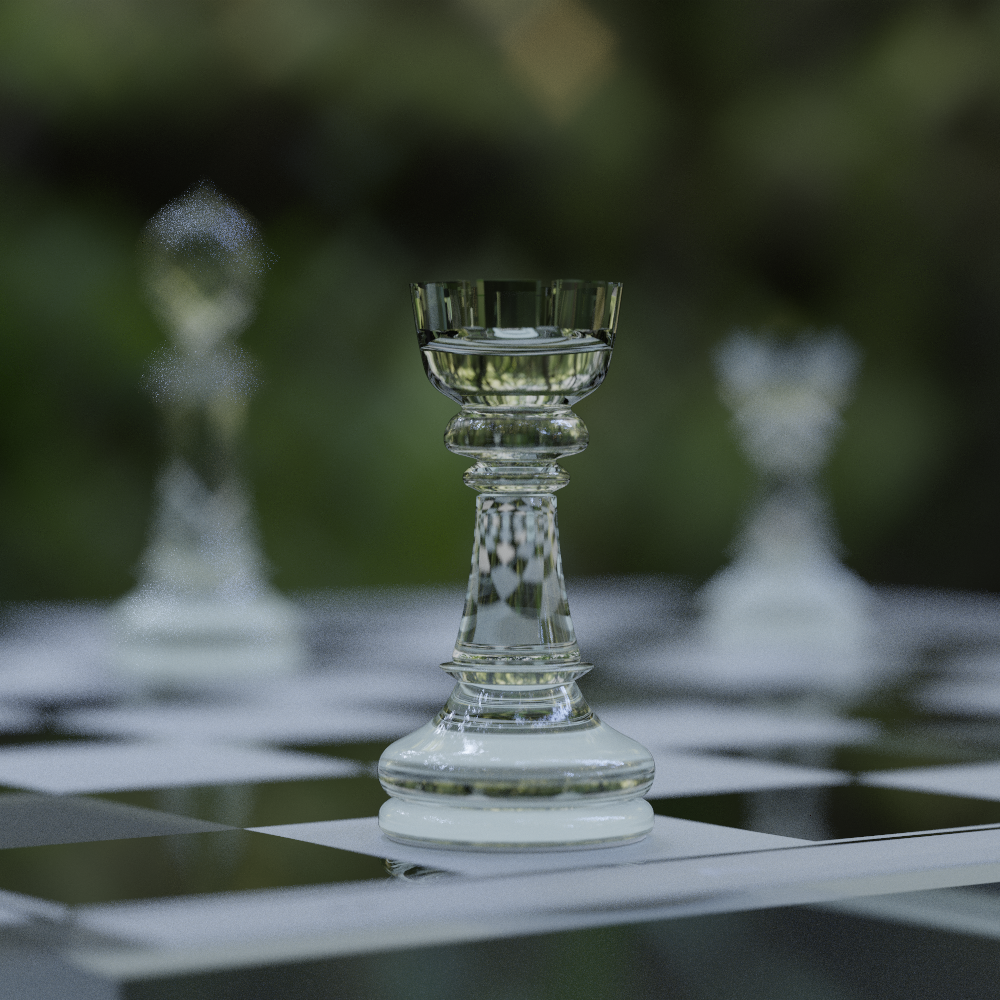}
    \end{tabular}%
    \caption{Examples of lighting effects that can be observed in the real world, produced by a ray-tracing-based renderer: soft shadows (top-left), color bleeding (top-right), reflection/refraction (bottom-left), and depth-of-field (bottom-right).}
  \label{fig:effects}
  \end{figure}

Ray tracing~\cite{whitted1980} is an essential and widely used technique in computer graphics. There are many applications using ray tracing, and one of the most popular and exciting use cases is photorealistic rendering. Please, take a look at Figure~\ref{fig:RT_or_photo}. Which picture do you think is a photograph? One of them is rendered by ray tracing (left), and the other is a photograph taken by a physical camera (right). The rendered image is practically indistinguishable from the photograph. This demonstrates how powerful ray tracing is. 
Light phenomena are described by the physical laws of optics, which are very complex in general. In computer graphics, we use a simplified model described by geometrical optics, modeling light propagation in terms of rays (i.e., straight lines). Even with this simplified model, we are able to render numerous lighting effects such as soft shadows, color bleeding, reflection/refraction, and depth-of-field (Figure~\ref{fig:effects}) in a single unified framework.

%
In this \articleitself{}, we introduce the basics of ray tracing and explain how to accelerate the computation of the rendering algorithm in HIP. We also show how to use a HIP ray tracing framework - HIPRT, leveraging hardware ray tracing features of AMD GPUs. We conclude this \articleitself{} with a list of references for further reading. 

%% file: sec/2_orochi.tex
\section{Orochi}
The natural way to implement a HIP application accelerated by the GPU is to use the HIP SDK or ROCm. A drawback of using vanilla HIP is that the host code is compiled either for AMD or NVIDIA backends, requiring recompilation each time we switch to a GPU of the other vendor. In this section, we introduce Orochi, an open-source library that allows to switch backends of different vendors at runtime \cite{amd_orochi}. Orochi is a wrapper built on top of the HIP API, loading HIP API functions directly from the driver libraries (\verb|dll| files on Windows or \verb|so| files on Linux), and thus it does not require the HIP SDK installed. 
Orochi API is designed such that developers who are familiar with HIP or CUDA APIs can use it easily. We use \verb|hip| prefix for the HIP API while Orochi uses \verb|oro| prefix for the Orochi API. What it does is simply redirect the function to the appropriate API underneath. Listing~\ref{lst:hip_sample} shows a sample code, initializing a GPU device and Orochi context. Listing~\ref{lst:orochi_sample} shows how this code is transformed to the Orochi API.



\FloatBarrier
\begin{lstlisting}[caption=A HIP sample code.,style=myHIPStyle2,label={lst:hip_sample}]
#include <hip/hip_runtime.h>
hipInit(0);
hipDevice device;
hipDeviceGet(&device, 0);
hipCtx ctx;
hipCtxCreate(&ctx, 0, device);
\end{lstlisting}
\FloatBarrier

\FloatBarrier
\begin{lstlisting}[caption=An Orochi sample code.,style=myHIPStyle2,label={lst:orochi_sample}]
#include <Orochi/Orochi.h>
oroInitialize(ORO_API_HIP, 0);
oroInit(0);
oroDevice device;
oroDeviceGet(&device, 0);
oroCtx ctx;
oroCtxCreate(&ctx, 0, device);
\end{lstlisting}
\FloatBarrier

Most of the code in the above example is the same except for the API prefix. Only addition to the code is line 2 in Listing \ref{lst:orochi_sample} of the program where we call \verb|oroInitialize()|. We pass \verb|ORO_API_HIP| to the function, which tells the library to load the HIP API from the driver libraries. If we want to use an NVIDIA GPU, we need to pass \verb|ORO_API_CUDA| instead. We can change the code path from HIP to CUDA without recompiling the application, allowing for dynamic switching between AMD GPU and NVIDIA GPU in runtime. We can also use both AMD and NVIDIA GPUs by initializing Orochi with both keys \verb+(ORO_API_HIP | ORO_API_CUDA)+. If the system has two GPUs, one AMD and one NVIDIA, Orochi loads both APIs, allowing us to use both devices at the same time in a single application. Listing~\ref{lst:devices} shows a sample code that loads the HIP and CUDA APIs on the first line, counts the number of AMD and NVIDIA devices, and prints the name and architecture of each GPU. The full source code of this and other examples can be found in the Orochi repository on GPUOpen\cite{amd_orochi}.

\begin{lstlisting}[caption=Querying device count and printing a device name and architecture of each device.,style=myHIPStyle2,label={lst:devices}]
oroInitialize(static_cast<oroApi>(ORO_API_CUDA | ORO_API_HIP), 0);
oroInit(0);
int nDevicesTotal;
oroGetDeviceCount(&nDevicesTotal);
int nAMDDevices;
oroGetDeviceCount(&nAMDDevices, ORO_API_HIP);
int nNVIDIADevices;
oroGetDeviceCount(&nNVIDIADevices, ORO_API_CUDA);
std::cout << "# of devices: " << nDevicesTotal << std::endl;
std::cout << "# of AMD devices: " << nAMDDevices << std::endl;
std::cout << "# of NV devices: " << nNVIDIADevices << std::endl;
for(int i = 0; i < nDevicesTotal; i++)
{
    oroDevice device;
    oroDeviceGet(&device, i);
    oroDeviceProp props;
    oroGetDeviceProperties(&props, device);
    std::cout << "Device name and architecture: " << props.name << " (" << props.gcnArchName << ")" << std::endl;
}
\end{lstlisting}

%% file: sec/3_rendering_triangles.tex
\section{Rendering Triangles}
We use rays as a straight line to mimic physical light behavior in the real world. As light bounces around multiple times until they absorbed in the scene's surface, we have to cast a lot of rays for the light simulation. Also, triangles are commonly used primitives to represent the geometry of the scene. So, we will start with rendering with triangles using ray casting in this section.

Recent AMD Radeon HIP-compatible GPUs since RDNA2 has the capability to accelerate the ray intersections; however, developers have to use device-specific low-level APIs and design the algorithm carefully to achieve maximum ray tracing performance.  
Thus, we generally recommend using the HIPRT SDK, which we introduce later in this \articleitself{}. Although using such a vendor-provided SDK is often the optimal choice for most user applications in terms of performance, numerical robustness, and ease of implementation, use of the SDK as a black box may end up with inefficient usage or make debugging harder. Thus, we first explain ray tracing without using these APIs so readers can understand the algorithms behind these APIs, so that we believe this helps readers to use the APIs properly, understand the limitations.

\subsection{Ray-Triangle Intersection}
\begin{figure}
\centering
\includegraphics[width=0.6\textwidth]{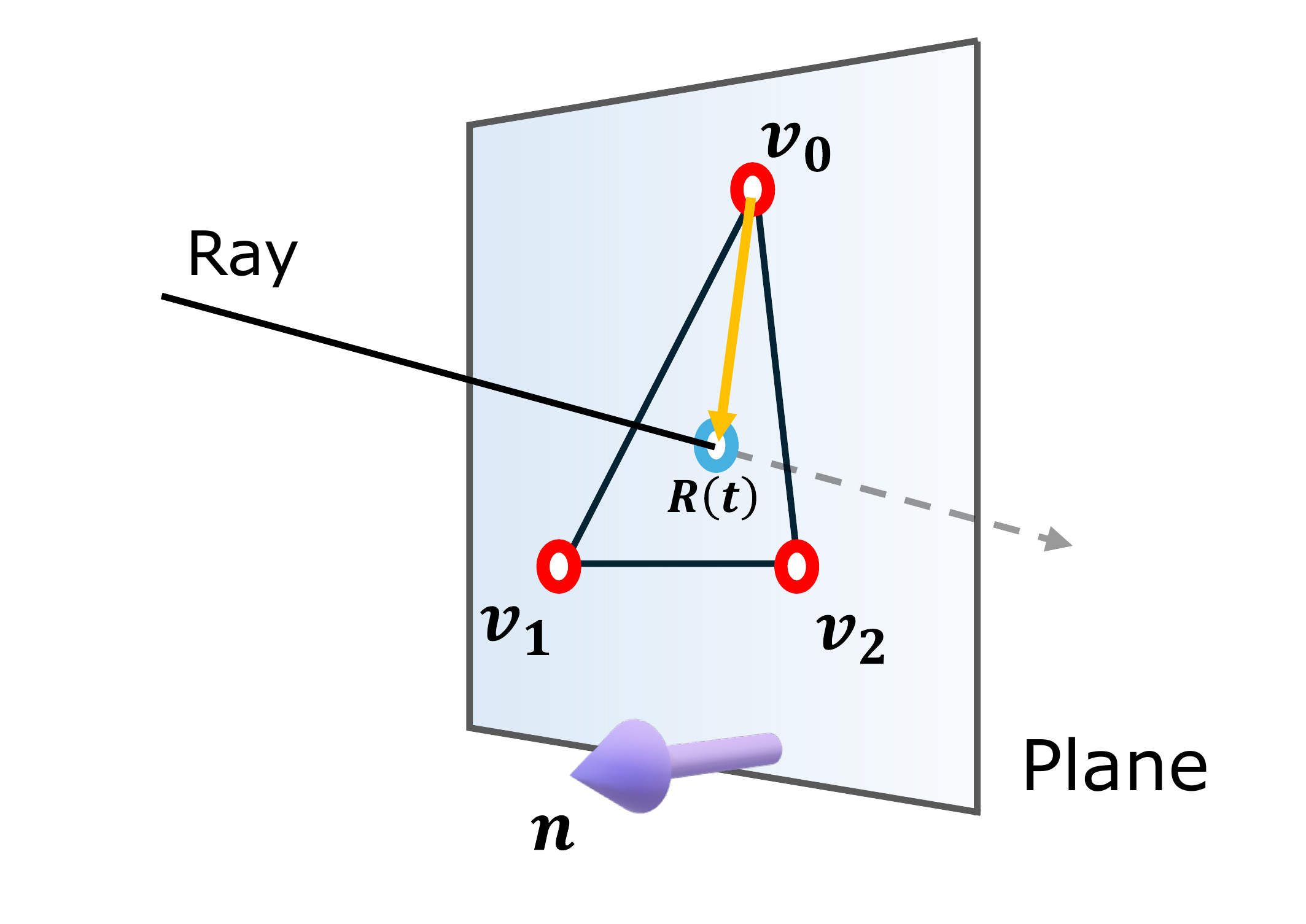}
\caption{To find the intersection with the plane of the triangle, we are looking for a value of t such $t$ that $\mathbf{R}(t) - \mathbf{v_0}$ is perpendicular to the normal of the triangle. }
\label{fig:rayplane}
\end{figure}

\begin{figure}
\centering
\includegraphics[width=\textwidth]{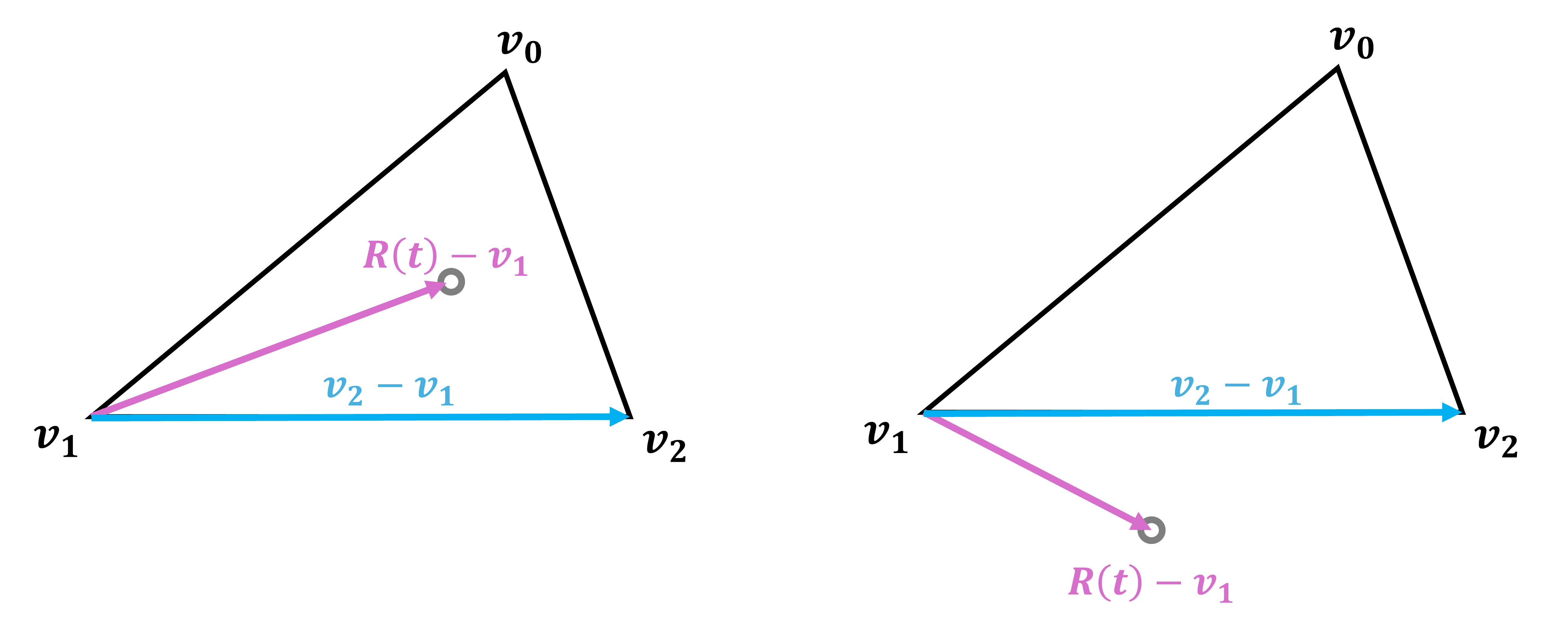}
\caption{ The left image shows that the hit point is inside with respect to the edge $ \mathbf{v_2}, \mathbf{v_1} $, which can be conditioned by $0 \leq \mathbf{n}\cdot((\mathbf{v_2}-\mathbf{v_1})\times(\mathbf{R}(t)-\mathbf{v_1})) $ while the right shows the hit point is outside against the edge. }
\label{fig:rayedge}
\end{figure}

The intersection of a ray and a triangle can be split into two steps. The first step is to find the intersection with a plane that the triangle belongs to, and the second step is to check whether the point is inside the triangle. Let us describe a ray as a parametric equation:
\begin{equation} 
\mathbf{R}(t) = \mathbf{o} + t\mathbf{d},
\label{eq:ray}
\end{equation}
where $t$ is a parameter (i.e., distance) of the ray, $\mathbf{o}$ is a ray origin, $\mathbf{d}$ is a ray direction. Given triangle vertices $(\mathbf{v_0}, \mathbf{v_1}, \mathbf{v_2})$, the normal of the triangle $\mathbf{n}$, we can describe the problem by the following equation:
\begin{equation} 
\mathbf{n}\cdot(\mathbf{R}(t) - \mathbf{v_0}) = 0.
\label{eq:plane}
\end{equation}
The intersection points is inside the triangle plane, and thus the vector $\mathbf{R}(t) - \mathbf{v_0}$ belongs to the plane as well. The dot product corresponds to a cosine of the angle between two vectors. If the the dot product is zero, two vectors are perpendicular. To satisfy the equality, point $\mathbf{R}(t)$ must lie on the plane, and thus it is the intersection point. Figure~\ref{fig:rayplane} illustrates the geometric interpretation of this equation. By plugging Equation~\ref{eq:ray} into Equation~\ref{eq:plane}, we can explicitly express parameter $t$ that satisfy the condition:
\begin{equation}
t = \frac{\mathbf{n}\cdot(\mathbf{v_0}-\mathbf{o})}{\mathbf{n}\cdot\mathbf{d}}.
\end{equation}
Since we already know the hit point $\mathbf{R}(t)$ in the plane, we can check if the point is inside the triangle. This test can be described as three edge tests as follows:
\begin{equation}
\begin{aligned}
&0 \leq \mathbf{n}\cdot((\mathbf{v_1}-\mathbf{v_0})\times(\mathbf{R}(t)-\mathbf{v_0})),\\
&0 \leq \mathbf{n}\cdot((\mathbf{v_2}-\mathbf{v_1})\times(\mathbf{R}(t)-\mathbf{v_1})),\\
&0 \leq \mathbf{n}\cdot((\mathbf{v_0}-\mathbf{v_2})\times(\mathbf{R}(t)-\mathbf{v_2})),
\end{aligned}
\end{equation}
where $\times$ denotes the cross product of three-dimensional vectors and $\mathbf{n}=(\mathbf{v_1} - \mathbf{v_0})\times (\mathbf{v_2} - \mathbf{v_1})$. Figure~\ref{fig:rayedge} shows one of the edge tests. Listing~\ref{lst:triangle_intersection} shows an implementation of the intersection test.

\begin{lstlisting}[caption=Ray-triangle intersection algorithm.,style=myHIPStyle2,label={lst:triangle_intersection}]
__device__ bool intersectRayTriangle(float& tOut, const float3& rayOrigin, const float3& rayDirection, const float3& v0, const float3& v1, const float3& v2)
{
    const float3 e0 = v1 - v0;
    const float3 e1 = v2 - v1;
    const float3 e2 = v0 - v2;
    const float3 n = cross(e0, e1);
    const float t = dot(v0 - rayOrigin, n) / dot(n, rayDirection);
    if (MIN_T <= t && t <= MAX_T)
    {
        const float3 p = rayOrigin + rayDirection * t;
        const float a = dot(n, cross(e0, p - v0));
        const float b = dot(n, cross(e1, p - v1));
        const float c = dot(n, cross(e2, p - v2));
        if (a < 0.0f || b < 0.0f || c < 0.0f)
            return false;
        tOut = t;
        return true;
    }
    return false;
}
\end{lstlisting}
\FloatBarrier

\subsection{Camera Model}
\begin{figure}
\centering
\includegraphics[width=\textwidth]{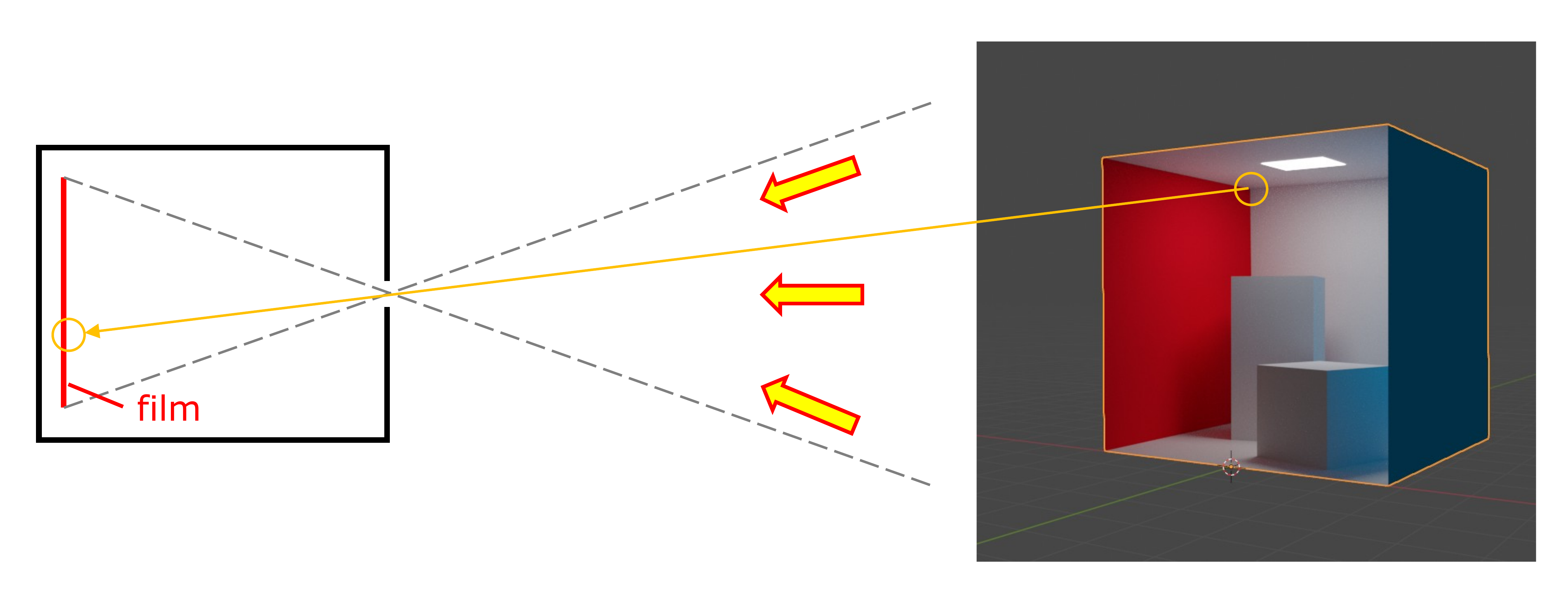}
\caption{Illustration of pinhole camera: light rays pass through the pinhole, projecting on the film on the opposite side.}
\label{fig:camera1}
\end{figure}

To render an image, we have to simulate how a real-world camera works. The simplest camera system is a pinhole camera: a box with a small hole on one of the sides and a film on the opposite side in the box. The small hole can work as a lens, such that the light coming from the outside is projected onto the film, forming an image. Figure~\ref{fig:camera1} illustrates how a pinhole camera works. 

\begin{figure}[h]
  \centering
  \begin{subfigure}[t]{0.48\linewidth}
    \centering
    \includegraphics[width=\textwidth]{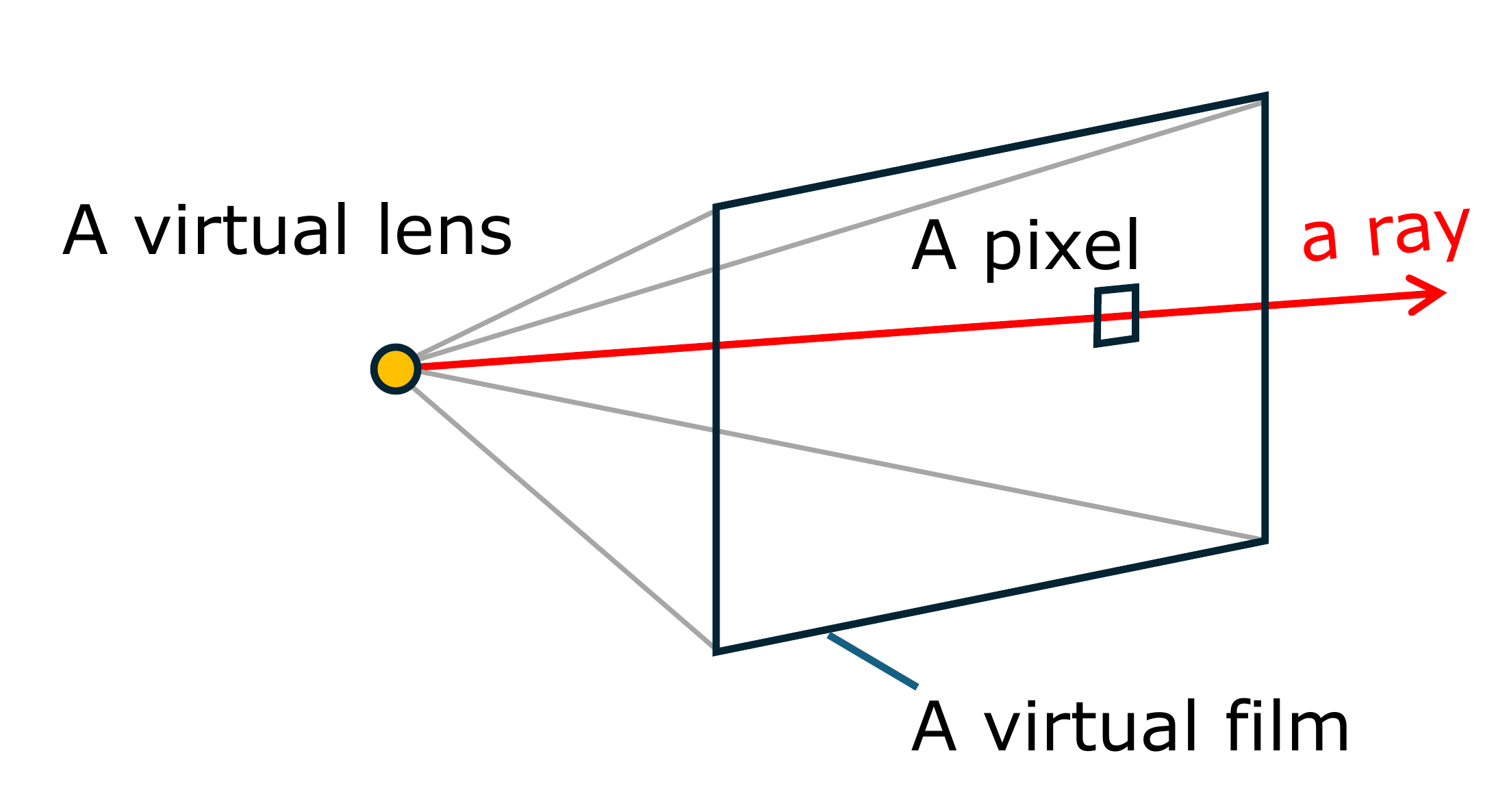}
    \caption{A simplified pinhole camera model}
    \label{fig:camera2}
  \end{subfigure}\hfill
  \begin{subfigure}[t]{0.48\linewidth}
    \centering
    \includegraphics[width=\textwidth]{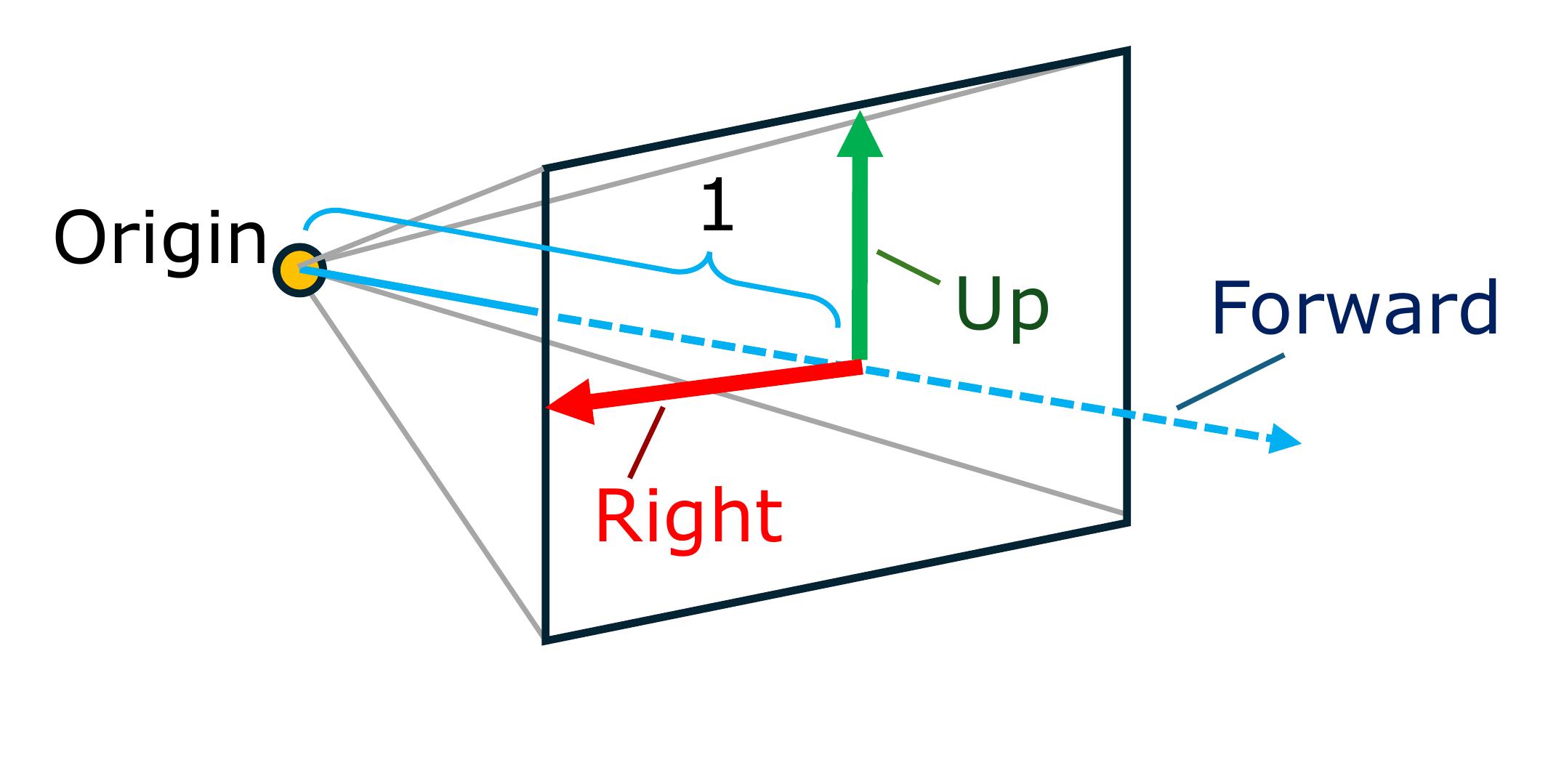}
    \caption{A parameterization of the camera} 
    \label{fig:camera3}
  \end{subfigure}
  \caption{Pinhole camera model commonly used in computer graphics.}
\end{figure}

In computer graphics, we often use an even simpler model of the real pinhole camera. A virtual film is placed in front of the lens, and the film is split into pixels (see Figure~\ref{fig:camera2}). In this model, we shoot a ray from the lens through a pixel into the virtual scene, and find an intersection with the ray in order to calculate the color of the pixel. We introduce a parameterization in Figure~\ref{fig:camera3}, where the lens location is as the origin, the size of the film as the up and right vectors, and the forward direction is implicitly represented as the cross product of the up and right vectors. Note that the distance from the origin and the virtual film is 1. Listing~\ref{lst:raygen} shows how to generate a camera ray given camera parameters. 

\FloatBarrier

\begin{lstlisting}[caption=Generating a camera (primary) ray passing throught a pixel on the virtual screen; u and v are normalized coordinate on the screen. , style=myHIPStyle2,label={lst:raygen}]
struct RayGenerator
{
    float3 m_origin;
    float3 m_right;
    float3 m_up;
    __device__ void getPrimaryRay(float3& ro, float3& rd, float u, float v)
    {
        float3 from = m_origin;
        float3 forward = normalize(cross(m_up, m_right));
        float3 to = m_origin + forward + lerp(-m_right, m_right, u) +
                    lerp(m_up, -m_up, v);
        ro = from;
        rd = normalize(to - from);
    }
};
\end{lstlisting}

\FloatBarrier

\subsection{Rendering Triangles}
 Each such ray passing through a pixel can be processed independently, and thus multiple rays can be processed in parallel, which makes it suitable to execute on the GPU. Listing~\ref{lst:one_triangle} shows a HIP implementation of rendering a single triangle.

\FloatBarrier
\begin{lstlisting}[caption=Rendering a single triangle using ray tracing.,style=myHIPStyle2,label={lst:one_triangle}]
__global__ void RenderTriangleKernel(uint8_t* pixels, RayGenerator rayGen, const int width, const int height, const float3 v0, const float3 v1, const float3 v2)
{
  const int tid = threadIdx.x + blockDim.x * blockIdx.x;
  if (width * height <= tid) { return; }
  const int xi = tid % width;
  const int yi = tid / width;
  float3 rayOrigin, rayDirection;
  rayGen.getPrimaryRay(rayOrigin, rayDirection, 
    static_cast<float>(xi) / width, static_cast<float>(yi) / height);
  const int pixelIdx = xi + (height - yi - 1) * width;
  float t;
  bool hit = intersectRayTriangle(t, rayOrigin, rayOrigin, v0, v1, v2);
  pixels[pixelIdx * 4 + 0] = (hit)?R_triangle:32;
  pixels[pixelIdx * 4 + 1] = (hit)?G_triangle:32;
  pixels[pixelIdx * 4 + 2] = (hit)?B_triangle:32;
  pixels[pixelIdx * 4 + 3] = 255;
}
\end{lstlisting}
\FloatBarrier

The code in Listing~\ref{lst:one_triangle} renders of a single triangle; nonetheless, scenes typically consist of many triangles. Since light hit the closest surface along the ray, we do not see objects behind of other (opaque) objects, and thus, we need to find the closest hit point among the triangles to simulate the light behavior. Listing~\ref{lst:many_triangles} shows an implementation of a na\"ive linear search, rendering multiple triangles. Figure~\ref{fig:cornellbox} shows an image rendered by ray tracing using the closes hit.

\FloatBarrier
\begin{lstlisting}[caption=Rendering multiple triangles using ray tracing.,style=myHIPStyle2,label={lst:many_triangles}]
  // Generate camera rays, etc.
  ...
  float minT = FLT_MAX;
  int triIdx = -1;
  for (int i = 0; i < triangles.size(); i++)
  {
    Triangle tri = triangles[i];
    float t;
    if (intersectRayTriangle(t, rayOrigin, 
      rayDirection, tri.vtx[0], tri.vtx[1], tri.vtx[2]))
    {
      if (t < minT)
      {
        triIdx = i; minT = t;
      }
    }
  }
  // Store the results.
  ...
}
\end{lstlisting}
\FloatBarrier


\subsection{Bonus: Lens Distortion}
\begin{figure}[h]
  \centering
  \begin{subfigure}[t]{0.48\linewidth}
    \centering
    \includegraphics[width=\textwidth]{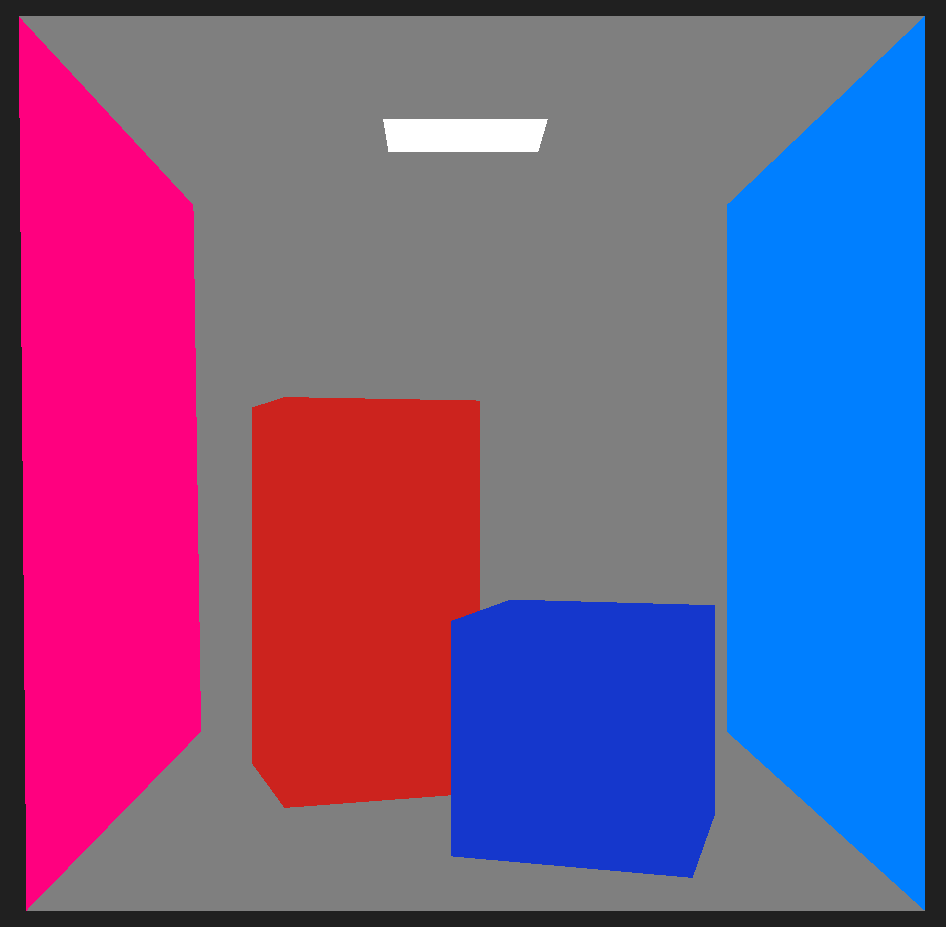}
    \caption{ A pinhole camera }
    \label{fig:cornellbox}
  \end{subfigure}\hfill
  \begin{subfigure}[t]{0.48\linewidth}
    \centering
    \includegraphics[width=\textwidth]{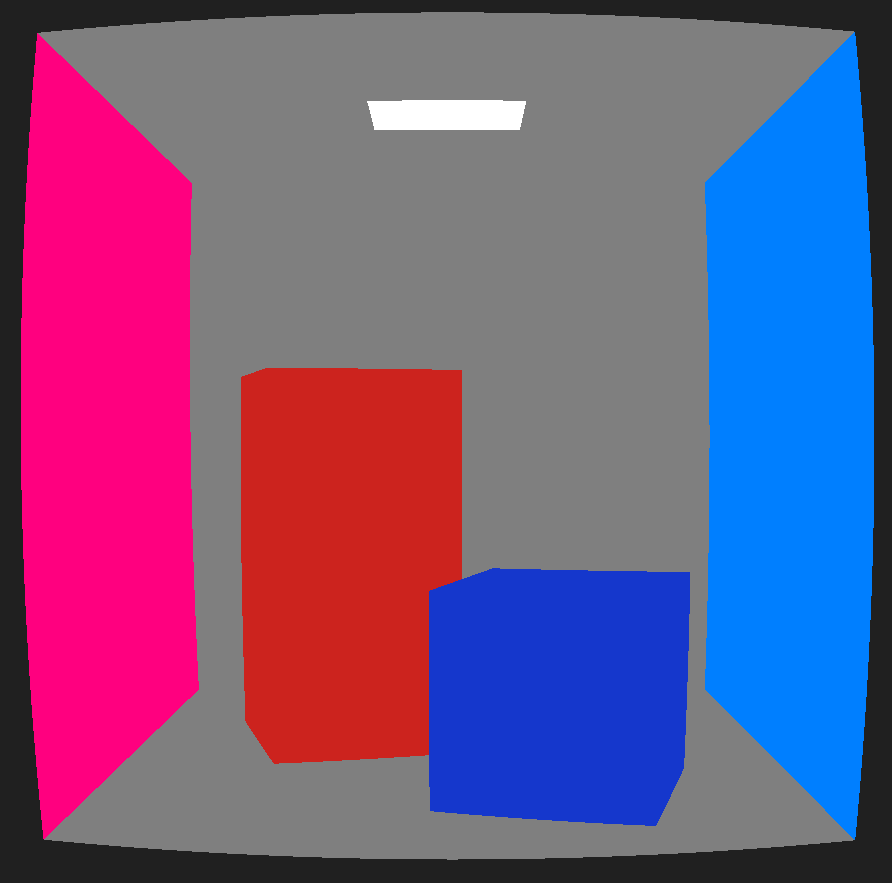}
    \caption{ A pinhole camera with radial distortion }
    \label{fig:cornellbox_d}
  \end{subfigure}
  \caption{Cornell box rendered by ray tracing using the pinhole camera model and the camera model with radial distortion.}
\end{figure}

A pinhole camera cannot model a real-world camera perfectly due to the shape and curvature of the lens, causing \emph{radial distortion} (e.g., straight lines are not perfectly straight). We can model a simple radial distortion as a mapping from a distorted location on the virtual film to a non-distorted space, where we can trace the corresponding ray as with a standard pinhole camera. We model the mapping as $ \textbf{p}_{non-distorted} = \frac{\textbf{p}_{distorted}}{1+c} $, where $c$ is a coefficient proportional to the squared distance from the center of the screen $c = d\|\textbf{p}_{distorted}\|^2$, where $d$ is a distortion parameter. The implementation is shown in Listing~\ref{lst:lens} and the rendered result is depicted in Figure~\ref{fig:cornellbox_d}. 



\FloatBarrier
\begin{lstlisting}[caption=Primary rays taking into account radial distortion of the lens., style=myHIPStyle2,label={lst:lens}]
struct RayGenerator
{
    float3 m_origin;
    float3 m_right;
    float3 m_up;
    __device__ void getPrimaryRay(float3& ro, float3& rd, float u, float v)
    {
        float3 from = m_origin;
        float3 forward = normalize(cross(m_up, m_right));

        float3 X = lerp(-m_right, m_right, u);
        float3 Y = lerp(m_up, -m_up, v);
        float3 p_distorted = X + Y;
        float r2 = dot(p_distorted, p_distorted);
        float d = -0.5f;
        float c = d * r2;
        float3 to = m_origin + forward + (X + Y) / (1.0f + c);
        ro = from;
        rd = normalize(to - from);
    }
};
\end{lstlisting}
\FloatBarrier

\FloatBarrier

%% file: sec/4_ambient_occlusion.tex
\section{Ambient Occlusion}
\begin{figure}
\includegraphics[width=\textwidth]{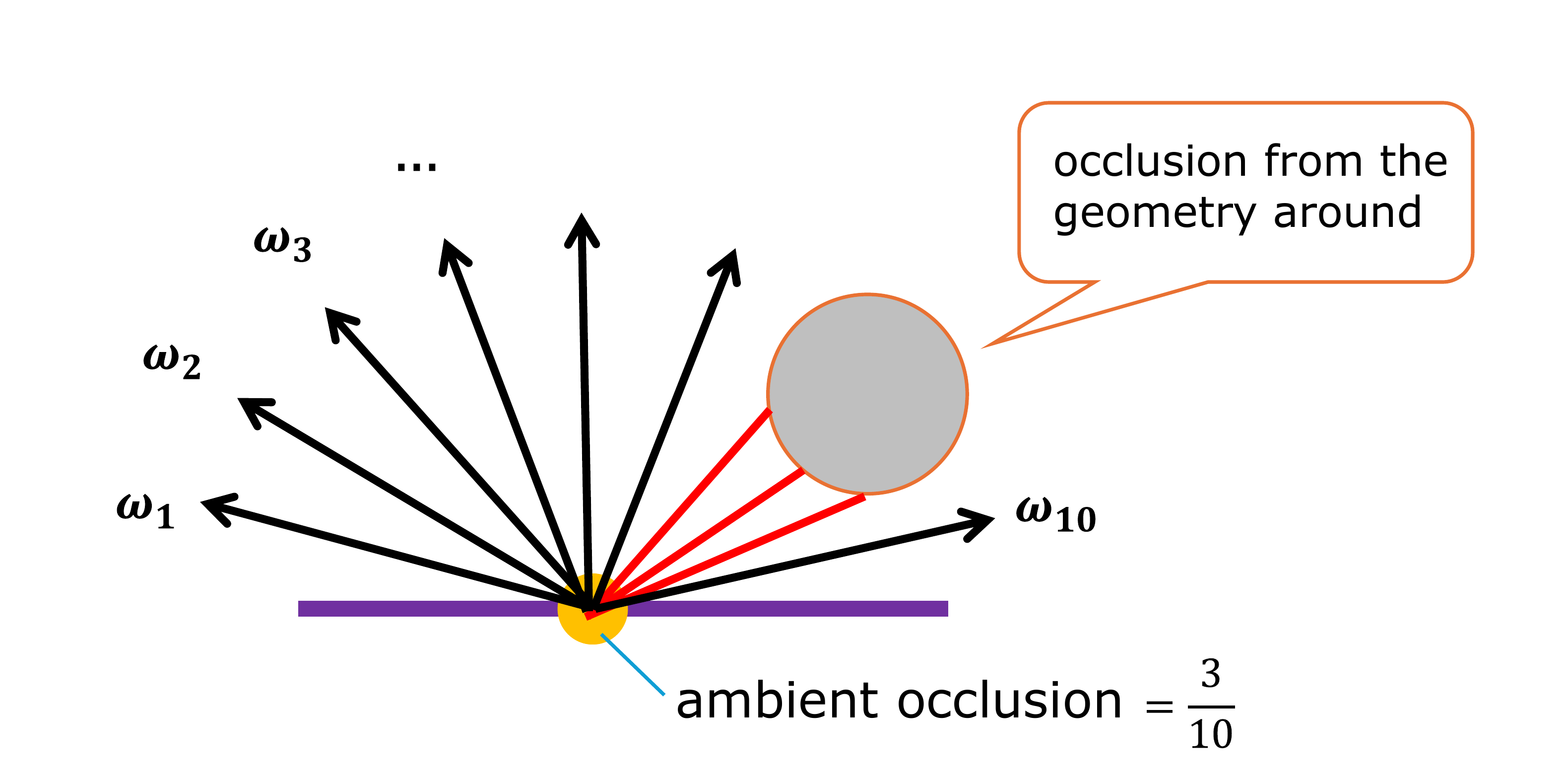}
\caption{An illustration how the incoming light is occluded by the geometry in the proximity of the shading point. }
\label{fig:ao}
\end{figure}

In the previous section, only the visibility from the camera is taken into account in the rendering. Since simulating full light transport is complex, let us start with a simple effect known as ambient occlusion, approximating indirect lighting by relative occlusion by geometry in a proximity of the shading point. Note that indirect lighting is lighting that does not come directly from a light source, reaching the shading points indirectly through surface interactions (e.g., refraction or transmission). The occlusion can be measured by shooting rays from the shading point. Figure~\ref{fig:ao} shows an example of ambient occlusion calculation with a limited number of rays. 


\subsection{Cosine Law and Cosine-Weighted Sampling}
\begin{figure}
\centering
\includegraphics[width=0.7\textwidth]{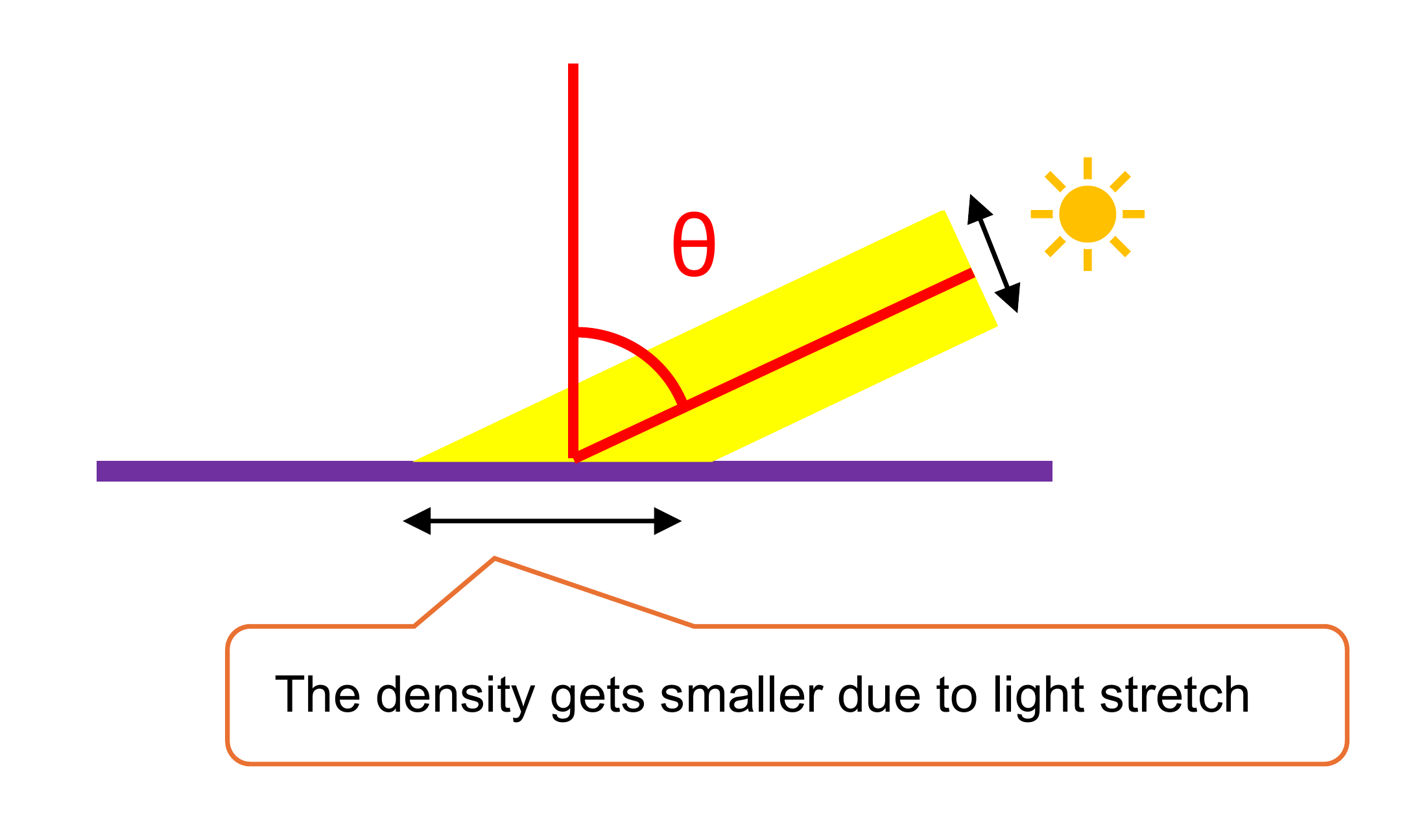}
\caption{An illustration of Lambert's cosine law: the incoming light is attenuated by the cosine term as the density of photons is smaller with larger angles.}
\label{fig:cos}
\end{figure}

In reality, incoming light does not illuminate the shading point equally from all directions. The amount of light reaching the shading point is proportional to the cosine of the angle between the direction and the normal. This is known as \emph{Lambert's cosine law}, which is illustrated in Figure~\ref{fig:cos}. To take this into account, we need not only to multiply the relative occlusion by the cosine, but also to normalize the result by $\pi$ such that a completely unoccluded shading point corresponds to $1$.

\begin{figure}
\centering
\includegraphics[width=1.0\textwidth]{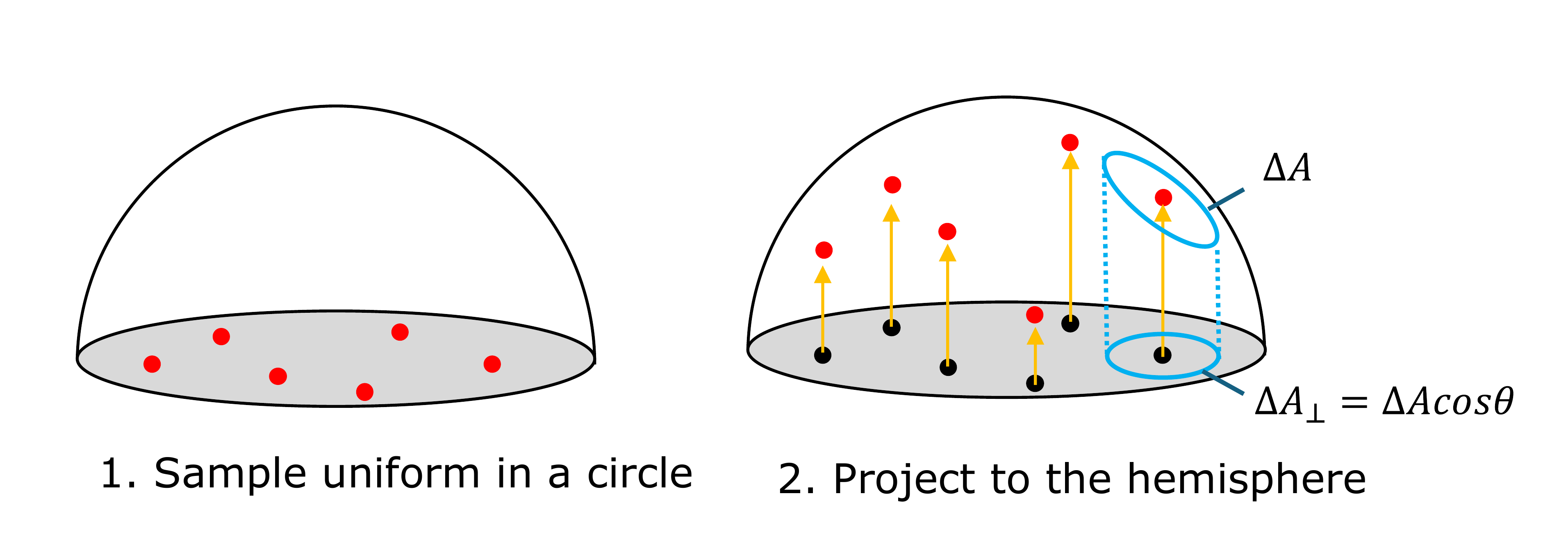}
\caption{An illustration of cosine-weighted sampling: we first uniformly sample points in the circle, and then project them to the hemisphere to get the random directions.}
\label{fig:cosSampling}
\end{figure}

A common practice is to sample directions randomly, unlike uniform intervals in Figure~\ref{fig:ao}. This is because such a regular sampling pattern produces visually disturbing correlated patterns in the rendered images. Due to the cosine term, it is beneficial to sample the direction according to a distribution proportional to the cosine of the angle instead of a uniformly distributed direction (we will discuss this in more detail in Section~\ref{sec:pt}). One of the simplest ways to realize this is to project a uniform distribution in a circle to a hemisphere. This works because of the area relationship between the circle and the hemisphere as shown in Figure~\ref{fig:cosSampling}, where the relationship between a small area on the hemisphere $ \Delta A$ and its projection $ \Delta A_{\perp}$ on the bottom circle can be described as $ \Delta A_{\perp} = \Delta A \cos \theta $. As the area on the circle is stretched by $ \frac{1}{\cos \theta} $ projecting to the hemisphere, the density of the distribution on the hemisphere is scaled by $ \cos \theta $. Changing the distribution using geometric relationships is often used to achieve the desired distribution (i.e., place samples proportional to the cosine value). For uniform distribution on a circle, we can use $ \phi = 2 \pi\xi_0, r = \sqrt{\xi_1} $ in the polar coordinate system, where $\xi_0, \xi_1$ are uniform random numbers in $[0,1]$, and $ x = r \cos\phi, z = r \sin\phi $. The square root here works as a cancellation of the radial compression introduced by the variable transformation from a unit square to a unit circle. Listing~\ref{lst:sampleHemisphere} shows an implementation of the cosine-weighted sampling. Since the hemisphere used in the implementation is centered around Y axis, we need to transform the the coordinates system of the shading point. We need to compute a basis of the coordinate system for the the transformation consisting of three vectors: normal, tangent, and bi-tangent. We have a normal vector, the tangent cane be pre-calculated by modeling software, but we can also use any of two vertices of the triangle, and the bi-tangent can be computed as a cross product of the normal and tangent vectors.




\subsection{Practical Consideration}
\begin{figure}[h]
  \centering
  \begin{subfigure}[t]{0.45\linewidth}
    \centering
    \includegraphics[width=\textwidth]{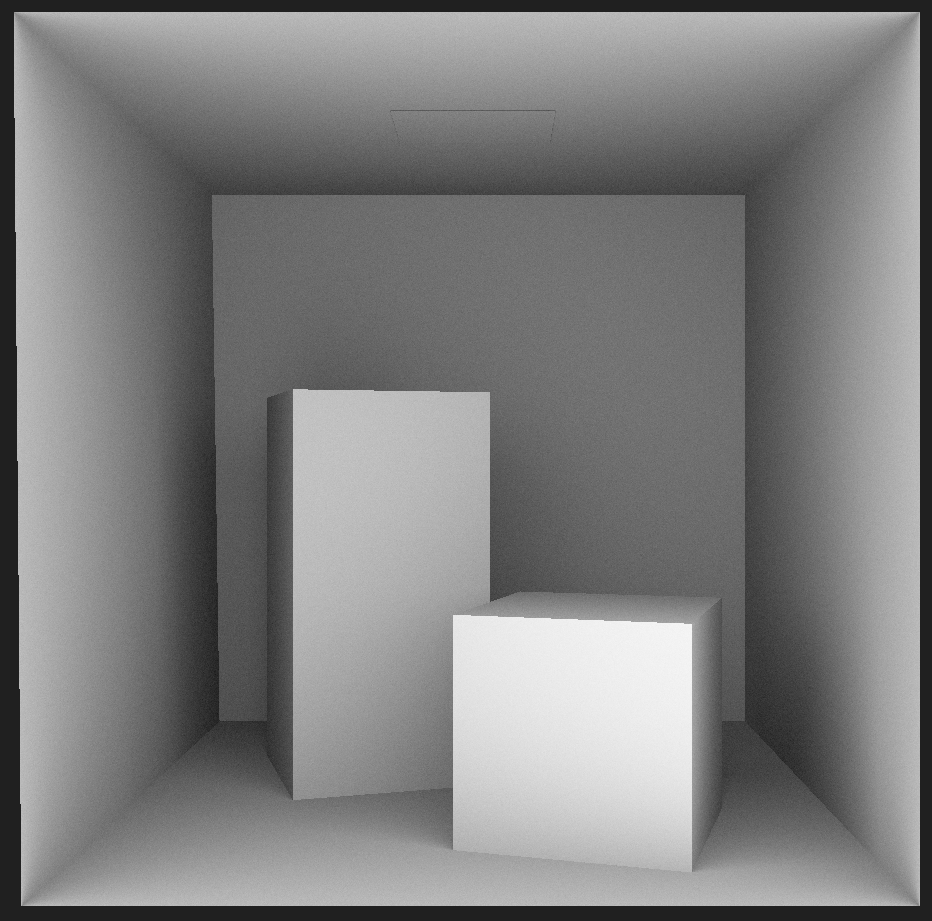}
\caption{ Ambient occlusion in grayscale }
\label{fig:ao_result}
  \end{subfigure}\hfill
  \begin{subfigure}[t]{0.45\linewidth}
    \centering
    \includegraphics[width=\textwidth]{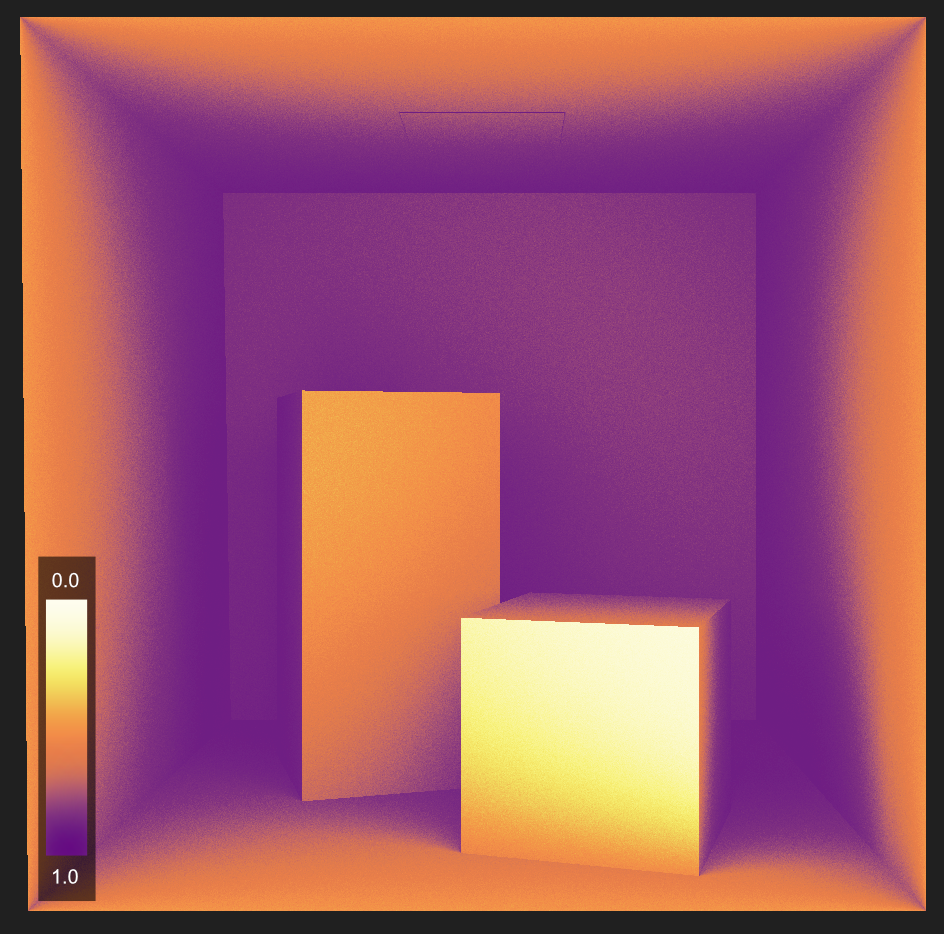}
    \caption{ A false-color rendering of ambient occlusion with a gradient look-up table. } 
    \label{fig:ao_mapped}
  \end{subfigure}
  \caption{Visualizations of Ambient Occlusion.}
\end{figure}

By using the cosine-weighted hemisphere sampling described above, Listing~\ref{lst:ao_code} shows the example code for the ambient occlusion calculation, and the result is shown in Figure~\ref{fig:ao_result}. Note that we add a small offset to the ray origin based on the normal at the point to avoid self-intersection with the base triangle. Ambient occlusion does not strictly follow physical light behavior and ignores multiple-bounce lighting effects, but it is far more interesting than the image we rendered in Figure~\ref{fig:cornellbox} and it resembles real-world occlusion. As it can be precomputed and baked into a texture, it has been used as a cheap approximation of shadows in real-time applications. In addition, ambient occlusion can be used for arbitrary non-photorealistic rendering. For example, Figure~\ref{fig:ao_mapped} shows a simple false color rendering by mapping the ambient occlusion value into a color table. 

\begin{figure}[h]
  \centering
  \begin{subfigure}[t]{0.3\linewidth}
    \centering
    \includegraphics[width=\textwidth]{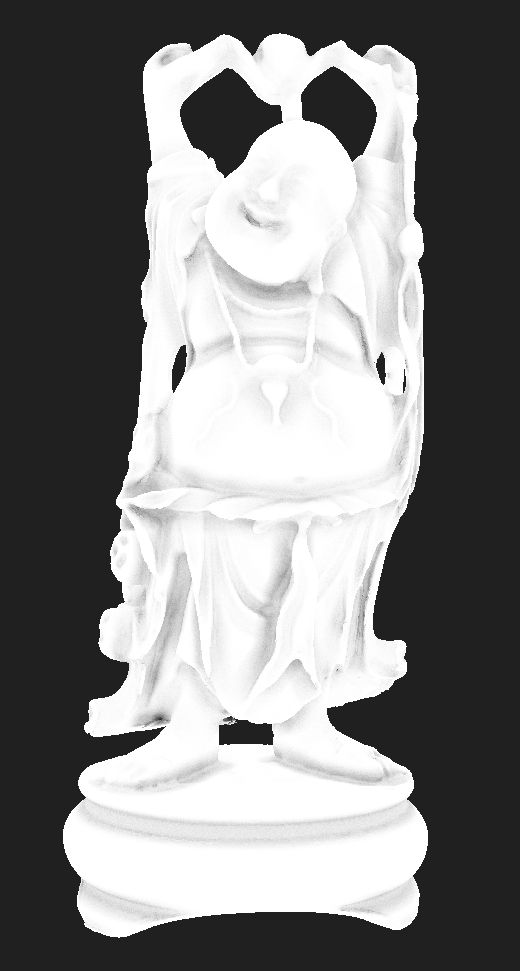}
    \caption{ 0.1 }
  \end{subfigure}\hfill
  \begin{subfigure}[t]{0.3\linewidth}
    \centering
    \includegraphics[width=\textwidth]{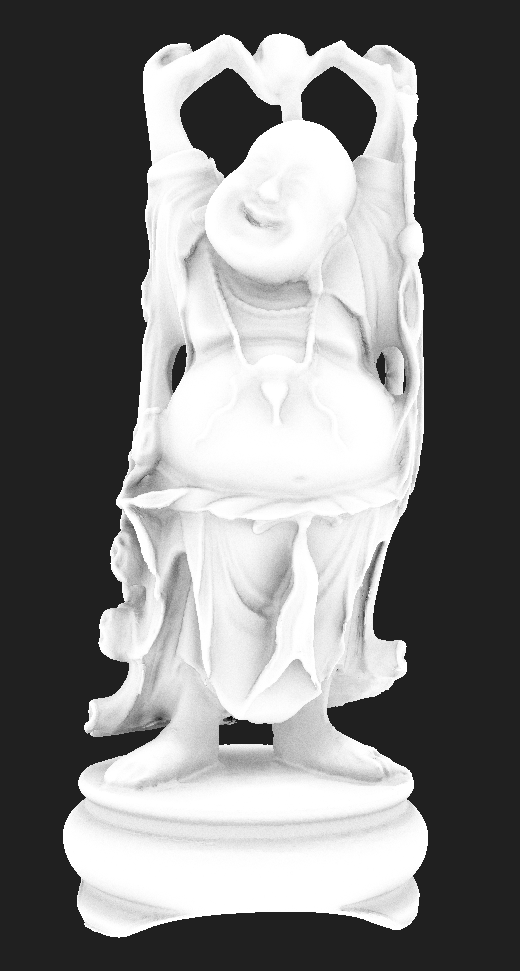}
    \caption{ 0.3 } 
  \end{subfigure}\hfill
  \begin{subfigure}[t]{0.3\linewidth}
    \centering
    \includegraphics[width=\textwidth]{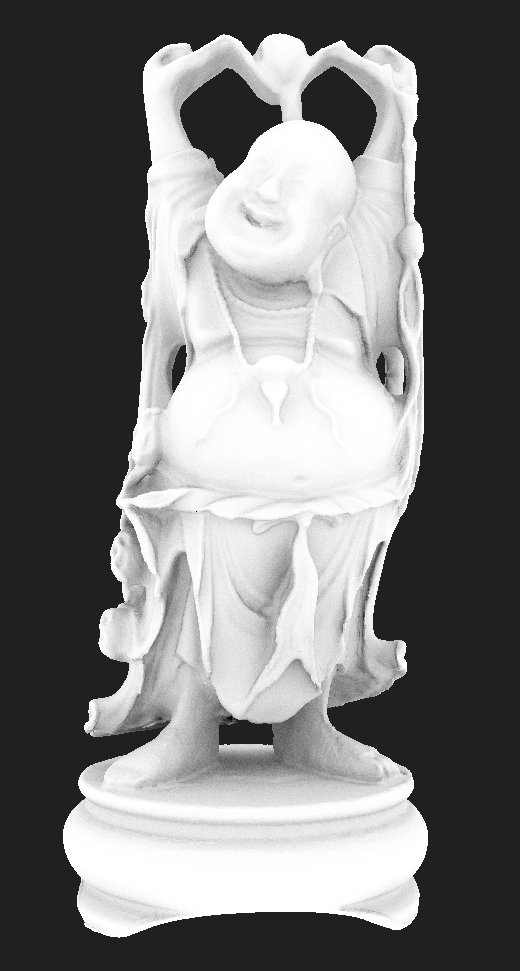}
    \caption{ 2.0 } 
  \end{subfigure}
  \caption{Ambient Occlusion with different ray lengths.}
  \label{fig:ao_length}
\end{figure}

We used ray tracing to find the closest hit along the ray, considering any intersection along the ray. In practise, we limit the length of a ray from the shading point to control the locality of the ambient occlusion. Thus, we do not need to test triangles behind the second point, and we can also stop immediately after finding first intersection (not necessarily the closest one) as we are interested only in occlusion. Ray tracing frameworks such as HIPRT support the maximum ray length and provide optimized any-hit kernels that can be used for such tests. Figure~\ref{fig:ao_length} shows the effect of the ray length in ambient occlusion. 


\FloatBarrier

\begin{lstlisting}[caption=Cosine-weighted hemisphere sampling.,style=myHIPStyle2,label={lst:sampleHemisphere}]
float3 sampleHemisphere(float xi_0, float xi_1)
{
  float phi = xi_0 * 2.0f * PI;
  float r = sqrtf(xi_1);
  // uniform in a circle
  float x = cosf(phi) * r;
  float z = sinf(phi) * r;
  // project to a hemisphere
  float y = sqrtf(fmax(1.0f - r * r, 0.0f));
  return {x, y, z};
}

\end{lstlisting}

\begin{lstlisting}[caption=Calculation of ambient occlusion. ,style=myHIPStyle2,label={lst:ao_code}]
const float3 p = rayOrigin + rayDirection * t;
const float3 aoRayOrigin = p + n * 0.0001f;
constexpr int N_RAYS = 1024;
int nOcclusions = 0;
for (int i = 0; i < N_RAYS; i++)
{
  float3 s = sampleHemisphere(random.uniformf(), random.uniformf());
  // The hemisphere is Y-up. Transform the Y axis to the normal
  float3 aoRayDirection = tangent0 * s.x + tangent1 * s.z + n * s.y;
  Intersection aoIsect;
  if (closesetHit(aoIsect, aoRayOrigin, aoRayDirection, triangles)) 
  {
    nOcclusions++;
  }
}
const float ao = static_cast<float>(nOcclusions) / N_RAYS;
\end{lstlisting}
\FloatBarrier


\FloatBarrier

%% file: sec/5_path_tracing.tex
\section{Path Tracing}
\label{sec:pt}
Path tracing~\cite{kajiya1986re} is one of the most popular techniques to render photorealistic images. It is used in a wide range of applications, from offline movie production to real-time rendering in games. Producing a photorealistic image means simulating how light behaves in the real world. In path tracing, we simulate the optical paths that come to the camera from light sources. This helps us to compute how much light arrives at each pixel in the final image. In the real world, there are infinitely many possible paths that light can take before it reaches the camera. Simulating all of them is computationally impossible.
\begin{figure}[h]
\centering
\includegraphics[width=1.0\textwidth]{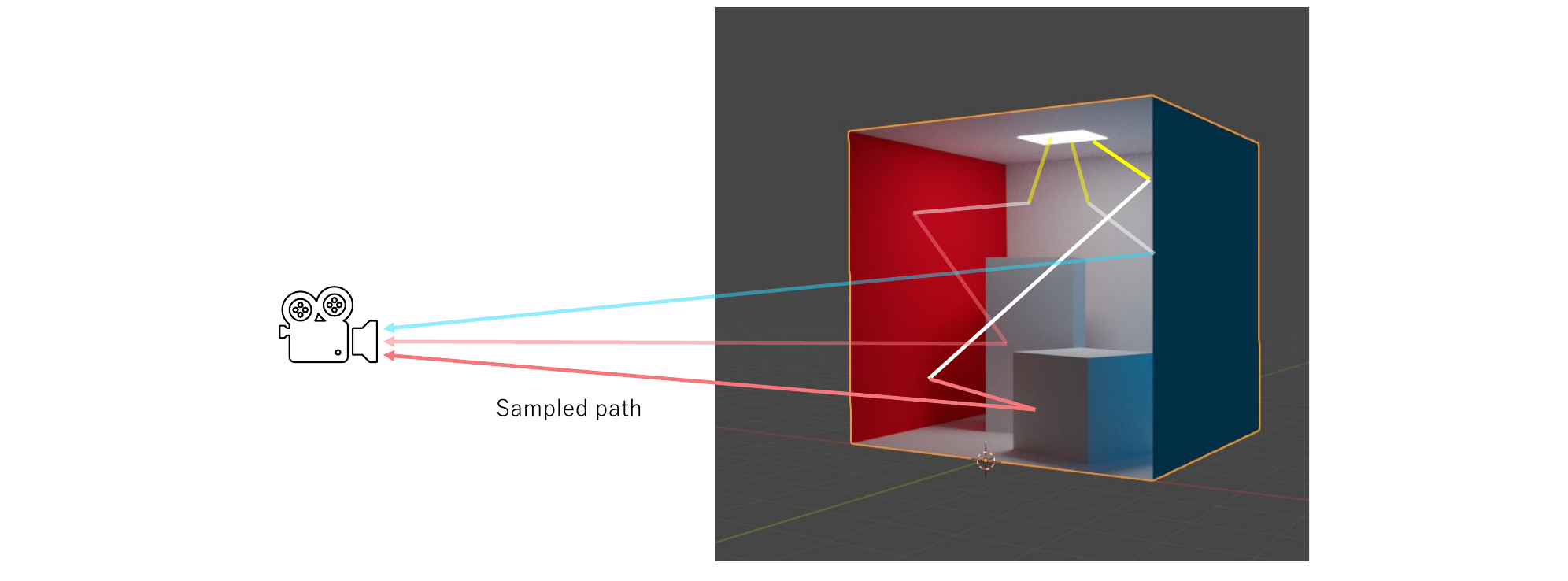}
\caption{A visualization of many light paths from a light source to the camera.}
\label{fig:light_paths}
\end{figure}

\begin{figure}[h]
\centering
\begin{subfigure}[t]{0.33\linewidth}
\centering
\includegraphics[width=\textwidth]{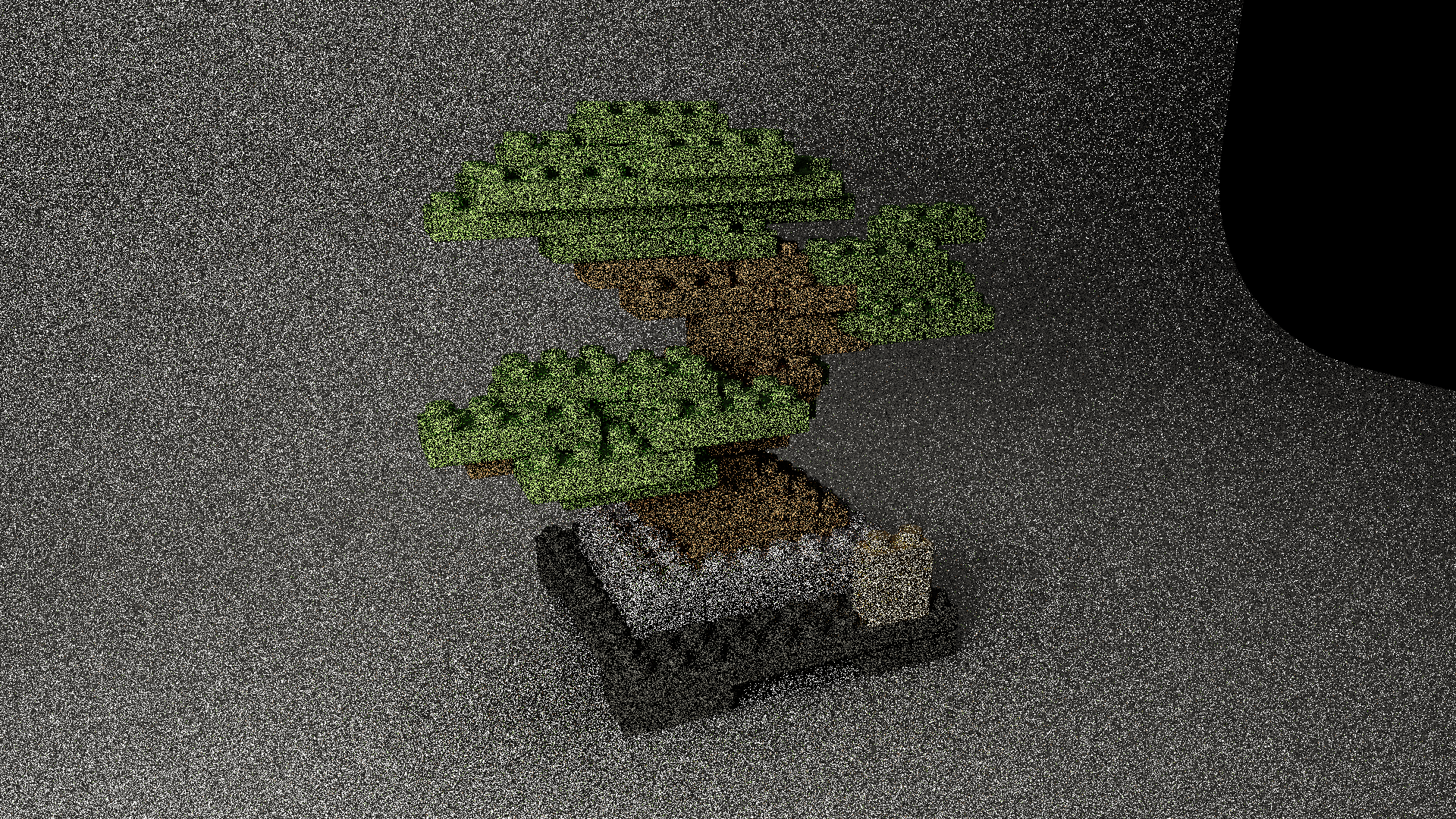}
\caption{16 samples}
\end{subfigure}\hfill
\begin{subfigure}[t]{0.33\linewidth}
\centering
\includegraphics[width=\textwidth]{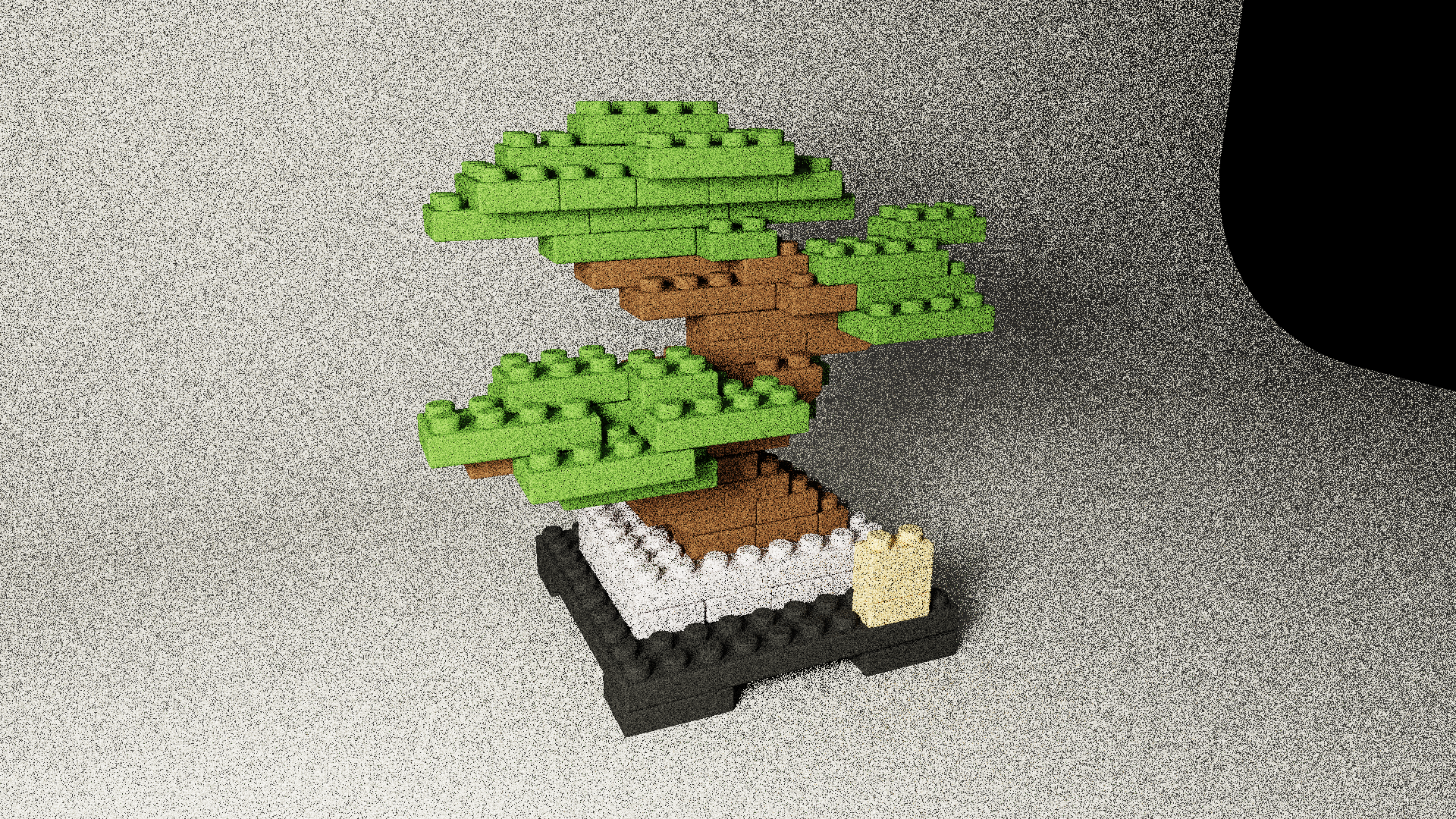}
\caption{128 samples} 
\end{subfigure}\hfill
\begin{subfigure}[t]{0.33\linewidth}
\centering
\includegraphics[width=\textwidth]{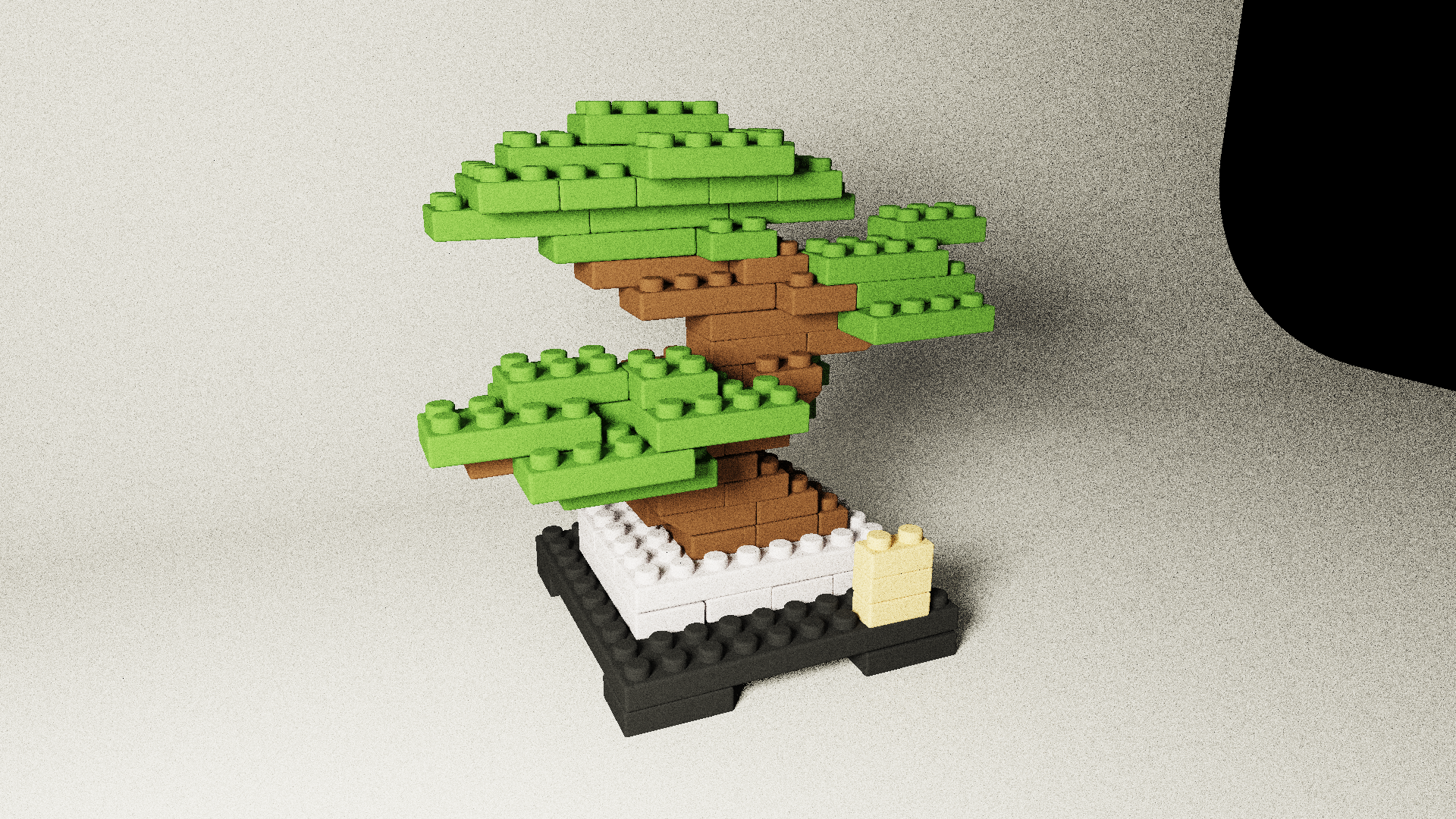}
\caption{1024 samples} 
\end{subfigure}
\caption{A comparison of images with different sample counts.}
\label{fig:pt_different_spp_compare}
\end{figure}

To address this, path tracing uses random sampling to generate a limited number of light paths (Figure~\ref{fig:light_paths}). This technique is called \emph{Monte Carlo ray tracing}. By averaging the results of many random paths, we can estimate the amount of light reaching each pixel. Ambient occlusion in the previous section is one example of Monte Carlo ray tracing. Because the method relies on random numbers, the resulting image contains noise. As we increase the number of samples (i.e., light paths), the noise decreases, and the image becomes clearer (see Figure~\ref{fig:pt_different_spp_compare}). In practical applications like movies and games, reducing noise while keeping the computation time short is a major challenge. Although we do not cover them in this section, many advanced techniques have been developed to tackle this challenge. \cite{kajiya1986re, veach1997thesis, jensen2001pm, bitterli2020restir}

\subsection{Algorithm}
In the real world, light travels from light sources to the camera. However, in path-tracing, we construct light paths in the opposite direction - from the camera toward the light sources. The reason for this is efficiency. If we trace paths from the light source, most of them would never reach the camera. Starting from the camera, we only trace paths that contribute to the image.

\begin{figure}[h]
\centering
\includegraphics[width=1.0\textwidth]{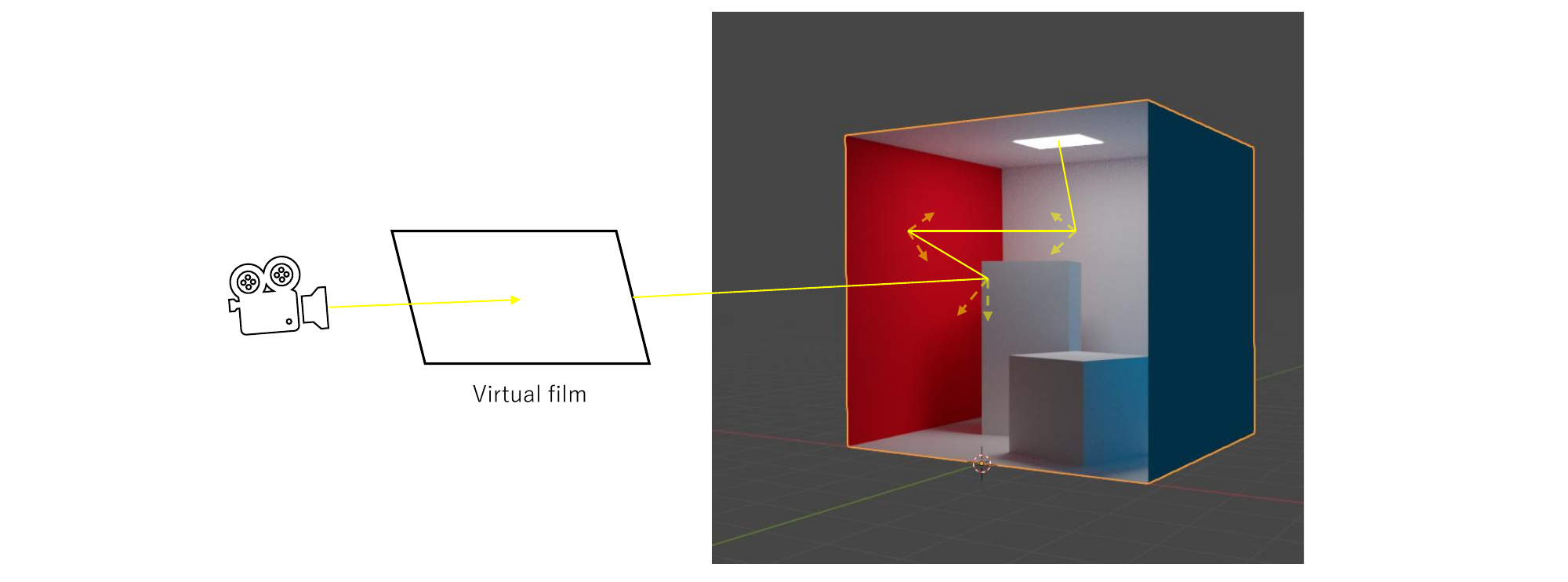}
\caption{Illustration of the path tracing algorithm.}
\label{fig:pt}
\end{figure}

We use ray-tracing to construct these paths. For each pixel, we first generate a ray that starts from the camera and passes through the pixel. We trace this ray into the scene and find its intersection with objects in the same manner as we did in the case of ambient occlusion. At the intersection point, we randomly sample a new direction and generate a new ray from that point. This process simulates reflection. We repeat this process, tracing the new ray, sampling a new direction, and so on, until the ray eventually hits a light source. When a ray reaches a light source, we add the light contribution to the pixel's color. Figure~\ref{fig:pt} illustrates this algorithm.

In the real world, when light hits a surface, some of the energy is absorbed, and the rest is reflected. To simulate this behavior, the path-tracing algorithm multiplies the reflectance of the surface by the light emission. The accumulated product of the reflectance along the ray is called \emph{throughput}. When a ray reaches a light source, we multiply the ray's throughput by the light emission and accumulate it in the pixel.

Since there can be multiple bounces before hitting a light source, we need to take all of them into account. We can denote the contribution for a single bounce case as $R_{0} \times E_{0}$, where $R_i$ is the reflectance, $E_i$ is the light emission, and the subscripts are the number of bounces. Also, we can write two-bounce case as $R_{0} \times R_{1} \times E_{1}$. Thus, we can sum them up to get the total contribution $L(i)$ for the number of bounces $i$ as follows:

$$
\begin{aligned}
L( i) =&  E_{0} + (R_{0} \times E_{1}) + (R_{0} \times R_{1} \times E_{2}) + \ldots + (R_{0} \times R_{1} \times ...\times R_{i-1} \times E_{i}).
\end{aligned}
$$

We can also simplify products of $R$ by throughput $T_i$, the accumulated product of the reflectance $T_i=R_0\times R_1 \times ...\times R_{i-1}$. Thus, we can simplify it as follows:

$$
\begin{aligned}
L( i) =  T_{0} \times E_{0} + T_{1} \times E_{1} +  T_{2} \times E_{2} + \ldots T_{i} \times E_{i} =  \sum _{k=0}^{i} T_{k} E_{k}.
\end{aligned}
$$

This lead us to a simple for loop to solve the light contribution as shown below.

\begin{lstlisting}[caption=A high-level Path tracing algorithm.,style=myHIPStyle2]
float3 radiance = {0.0f, 0.0f, 0.0f};
float3 throughput = {1.0f, 1.0f, 1.0f};
for (int depth = 0; depth < options.maxDepth ; ++depth)
{
    radiance += throughput * emissive(depth);
    throughput *= reflectance(depth);
}
\end{lstlisting}

Listing~\ref{lst:pt_code} shows a complete example code of this algorithm, including ray tracing. Note that the idea and the structure of the algorithm are the same as the code above.

\FloatBarrier
\begin{lstlisting}[caption=Path tracing algorithm.,style=myHIPStyle2,label={lst:pt_code}]
Ray ray = makeRay(ro, rd);
float3 radiance = {0.0f, 0.0f, 0.0f};
float3 throughput = {1.0f, 1.0f, 1.0f};
for (int depth = 0; depth < options.maxDepth; ++depth)
{
  Intersection isect;
  if (!raytrace(ray, hiprtGeom, isect))
  {
    // hit nothing
    radiance += throughput * options.skyColor;
    break;
  }
  Triangle hitTriangle = triangles[isect.index];
  if (hasEmission(hitTriangle))
  {
    // hit light source
    radiance += throughput * triangles[isect.index].emissive;
    break;
  }
  // Evaluate hit position and hit normal 
  SurfaceInfo surf = makeSurfaceInfo(ray, isect, triangles);
  float3 wo;
  {
    TangentBasis basis =
      makeTangentBasis(surf.n, isect.index, triangles);
    // Sample next ray direction in tangent space
    float3 woLocal = sampleHemisphere(
      random.uniformf(), random.uniformf());  
    // Transform direction from tangent space to world space
    wo = localToWorld(wo_local, basis);
  }
  // Update throughput
  throughput *= hitTriangle.color;
  // Generate next ray
  ray = makeRay(offsetRayPosition(surf.p, surf.n), wo);
}

// Write results to the accumulation buffer
accumulation[pixelIdx] += {radiance.x, radiance.y, radiance.z, 1.0f};
\end{lstlisting}
\FloatBarrier

Figure~\ref{fig:pt_different_spp_compare} shows rendering results using path tracing with different sample counts. Fewer samples result in noisier images. The amount of noise depends on how likely the randomly generated light paths are to reach a light source. If most of the sampled paths miss the light source, the estimation becomes highly variable, leading to visible noise. To reduce noise, it is important to sample light paths that are more likely to reach the light source. This idea is known as \emph{importance sampling}. In the next section, we introduce one such importance sampling technique called \emph{Next event estimation}.

\subsection{Next Event Estimation}
\begin{figure}
\centering
\includegraphics[width=0.8\textwidth]{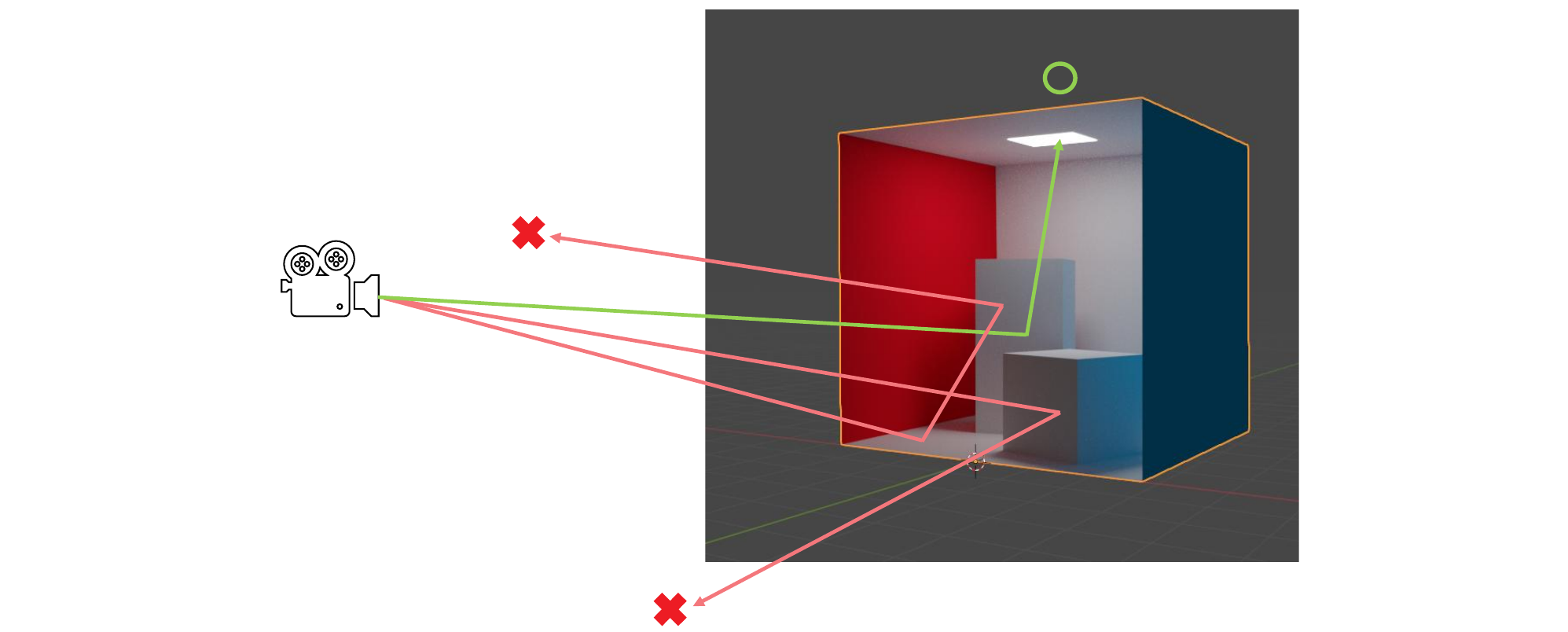}
\caption{Failure cases of basic path tracing.}
\label{fig:pt_problem}
\end{figure}

One of the main issues with basic path tracing is that rays are sampled randomly without considering the positions of light sources. As a result, many sampled paths fail to reach any light source, leading to an inefficient computation which results in high variance in the final image (Figure~\ref{fig:pt_problem}). Next event estimation~\cite{kajiya1986re, veach1997thesis} addresses this problem by explicitly taking the positions of light sources into account during sampling.

\begin{figure}[t]
\centering
\includegraphics[width=0.8\textwidth]{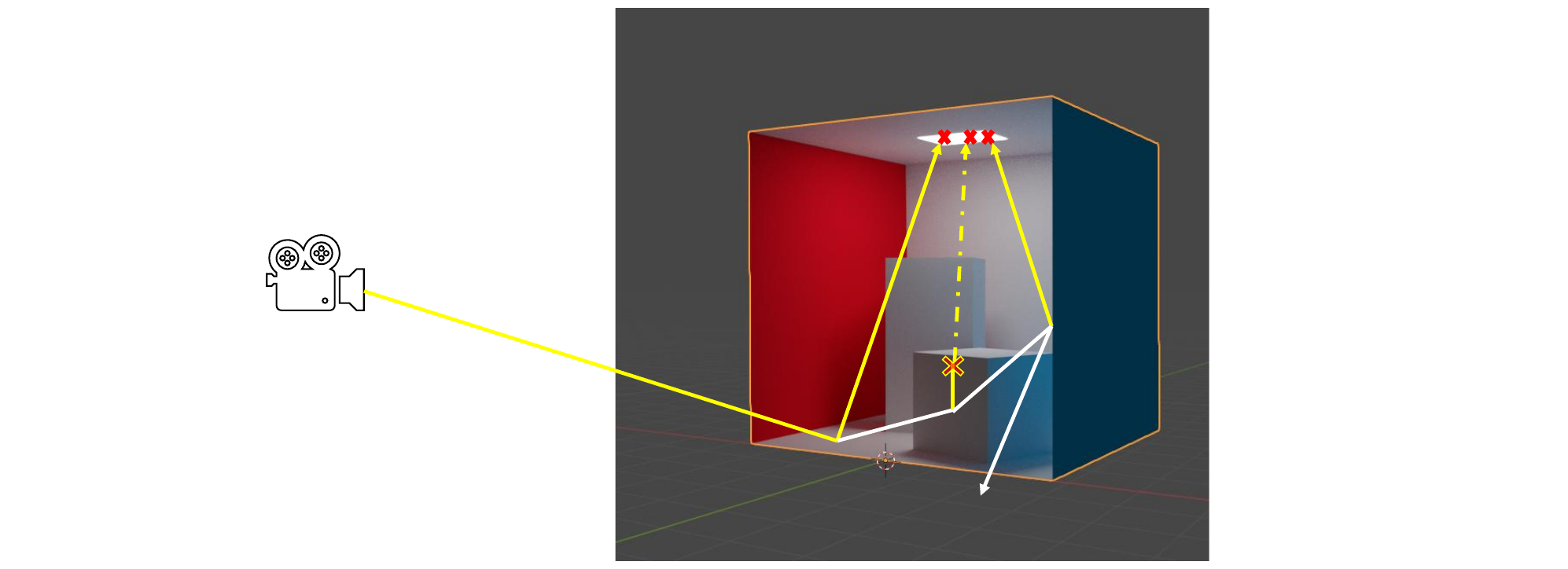}
\caption{Next event estimation}
\label{fig:nee_idea}
\end{figure}

At each surface intersection, next event estimation randomly selects one of the light sources in the scene. Once a light source is selected, a point on the light source is sampled. Then, a ray is generated to connect the intersection point and the sampled point on the light source. This ray is called a \emph{shadow ray}. If a shadow ray is not blocked by any objects, we accumulate the light contribution to the pixel. Whether a shadow ray is blocked or not is determined by ray tracing (similarly to ambient occlusion). This is the basic idea behind the next event estimation algorithm. Figure~\ref{fig:nee_idea} illustrates this idea. Listing~\ref{lst:nee} is an example code of the next event estimation algorithm.

\FloatBarrier
\begin{lstlisting}[caption=Next event estimation algorithm.,style=myHIPStyle2,label={lst:nee}]
// ...
for (int depth = 0; depth < options.maxDepth; ++depth)
{
  // raytrace
  // ...
  Triangle hitTriangle = triangles[isect.index];
  if (hasEmission(hitTriangle))
  {
    // hit light source
    if (depth == 0) {radiance += throughput * hitTriangle.emissive;}
    break;
  }
  // Evaluate hit position and hit normal 
  // ...
  // Sample a position on the light source
  LightSample lightSample =
    sampleLight(triangles, lights, random.uniformf(),
                   random.uniformf(), random.uniformf());
  // Create a path between the hit position and the light source
  {
     Triangle lightTriangle = triangles[light_sample.index];
     const float V =
       checkVisibility(surf.p, surf.n, lightSample.p, hiprtGeom);
     const float3 brdf = 1.0f / PI * hitTriangle.color;
     const float G =
       geometryTerm(surf.p, surf.n, lightSample.p, lightSample.n);
     const float lightPdf =
       1.0f / lights.size() * 1.0f / calculateArea(lightTriangle);
     radiance +=
       throughput * brdf * G * V * lightTriangle.emissive / lightPdf;
  }
  // Sample next ray direction
  // ...
  // Update throughput
  // ...
  // Generate next ray
  // ...
}

// Write results to the accumulation buffer
// ...
\end{lstlisting}
\FloatBarrier

\begin{figure}[h]
\centering
\begin{subfigure}[t]{0.5\linewidth}
\centering
\includegraphics[width=\textwidth]{sec/figs/pt/pt_16spp.png}
\caption{Path tracing}
\end{subfigure}\hfill
\begin{subfigure}[t]{0.5\linewidth}
\centering
\includegraphics[width=\textwidth]{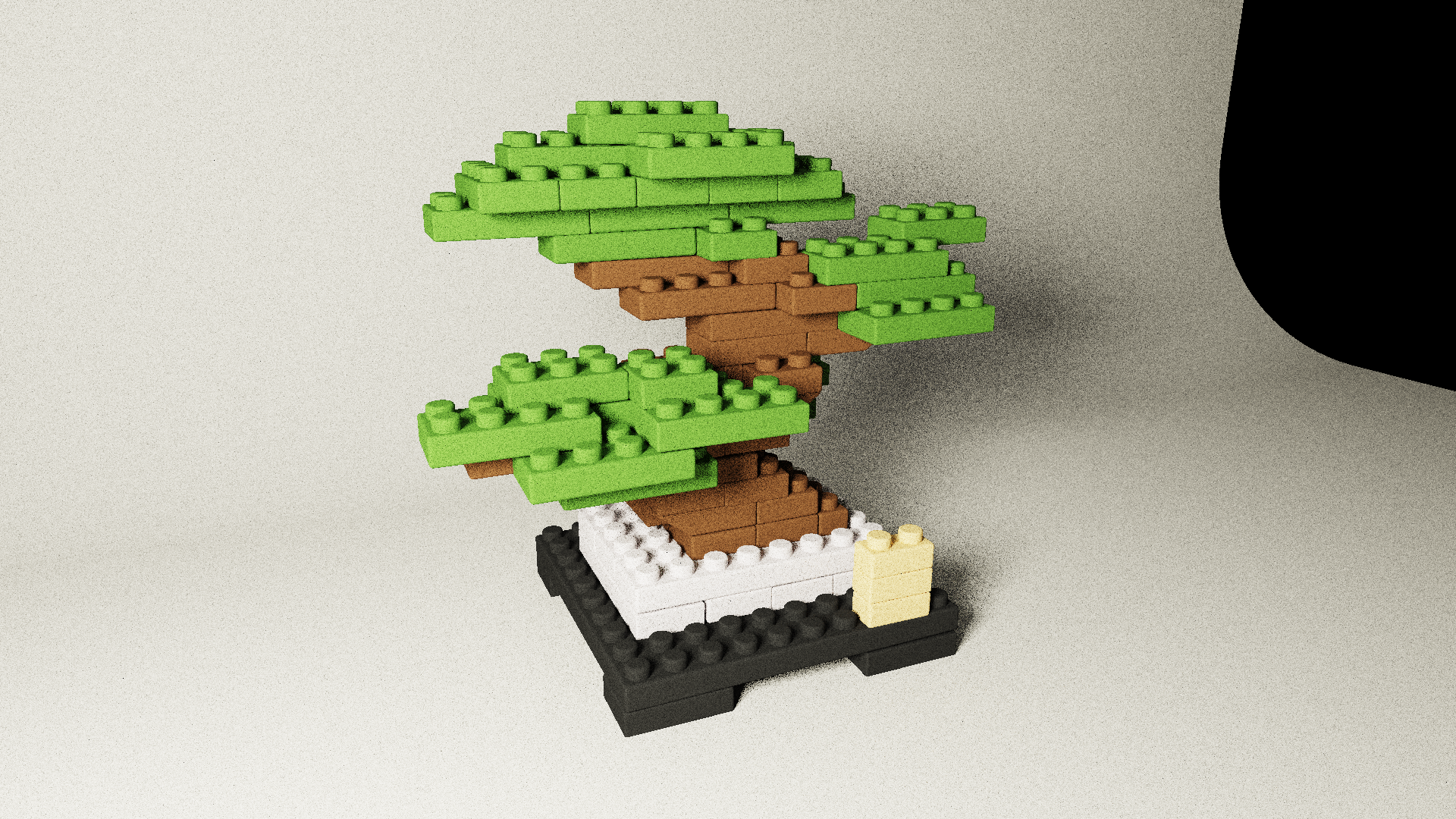}
\caption{Next event estimation} 
\end{subfigure}\hfill
\caption{A comparison of Path tracing and Next event estimation in 16 samples.}
\label{fig:pt_nee_compare}
\end{figure}

The next event estimation algorithm is quite similar to basic path tracing, but there are a few important differences. First, in Next event estimation, the contribution from rays that hit a light source directly is not added — except when the ray is at depth 0. This is to avoid counting the same light path twice. Second, the way we compute the contribution from the connection between the surface point and the sampled point on the light is different. As shown in the code example, several terms are multiplied with the light intensity, including the \emph{bidirectional reflectance distribution function (BRDF)}, \emph{geometry term}, and \emph{probability density function (PDF)} of the sampled ray. These terms are essential to make the estimation statistically unbiased. If we omit them, the rendered result will differ from basic path tracing. Although detailed explanations of these terms are beyond the scope of this book, we encourage interested readers to read \cite{pharr2023} for more information.
Figure~\ref{fig:pt_nee_compare} shows rendering results of Next event estimation. Compared to basic path tracing with the same number of samples, the noise is significantly reduced. As this example demonstrates, designing better methods to sample light paths is crucial for reducing noise in the rendered image.

%% file: sec/6_bvh.tex
\section{Bounding Volume Hierarchy (BVH)}
In the previous sections, we tested all triangles sequentially (i.e., linear time complexity) to find the intersections with a given ray. In practice, we deal with scenes of significantly higher complexity, consisting of millions of triangles (or other geometric primitives). Testing all triangles in a linear fashion becomes practically infeasible even on highly parallel GPUs already for scenes of moderate complexity. Therefore, significant effort has been devoted to reducing the number of intersection tests. The core idea is to arrange the triangles into spatial data structures that exploit the spatial proximity of the triangles such that we can efficiently cull parts of the scene (e.g., a subset of triangles) that are certainly not intersected. In this section, we focus on the bounding volume hierarchy (BVH), one of the most widely adopted acceleration data structures in modern ray tracing frameworks.

\subsection{Bounding Volumes} 
The idea is to enclose scene objects (or any subset of triangles) within simpler bounding volumes that can be easily tested for an intersection. If a ray does not intersect the bounding volume, we know that there is no intersection with the objects inside, and thus we do not need to test the objects inside. There are various types of bounding volumes, typically balancing culling efficiency (how tight the bounding volume is) and intersection efficiency (how quickly we can compute the intersection). Tighter bounding volumes can better cull rays thanks to reduced empty space, but the intersection test is typically more complex. For ray tracing, we use axis-aligned bounding boxes (AABBs), defined by minimum and maximum points. For elongated diagonal triangles, it may be beneficial to use oriented bounding boxes (represented as AABBs with additional rotation matrices) (OBBs), which better fit these non-axis aligned shape.

\begin{figure}
    \centering
    \includegraphics[width=1.05\linewidth]{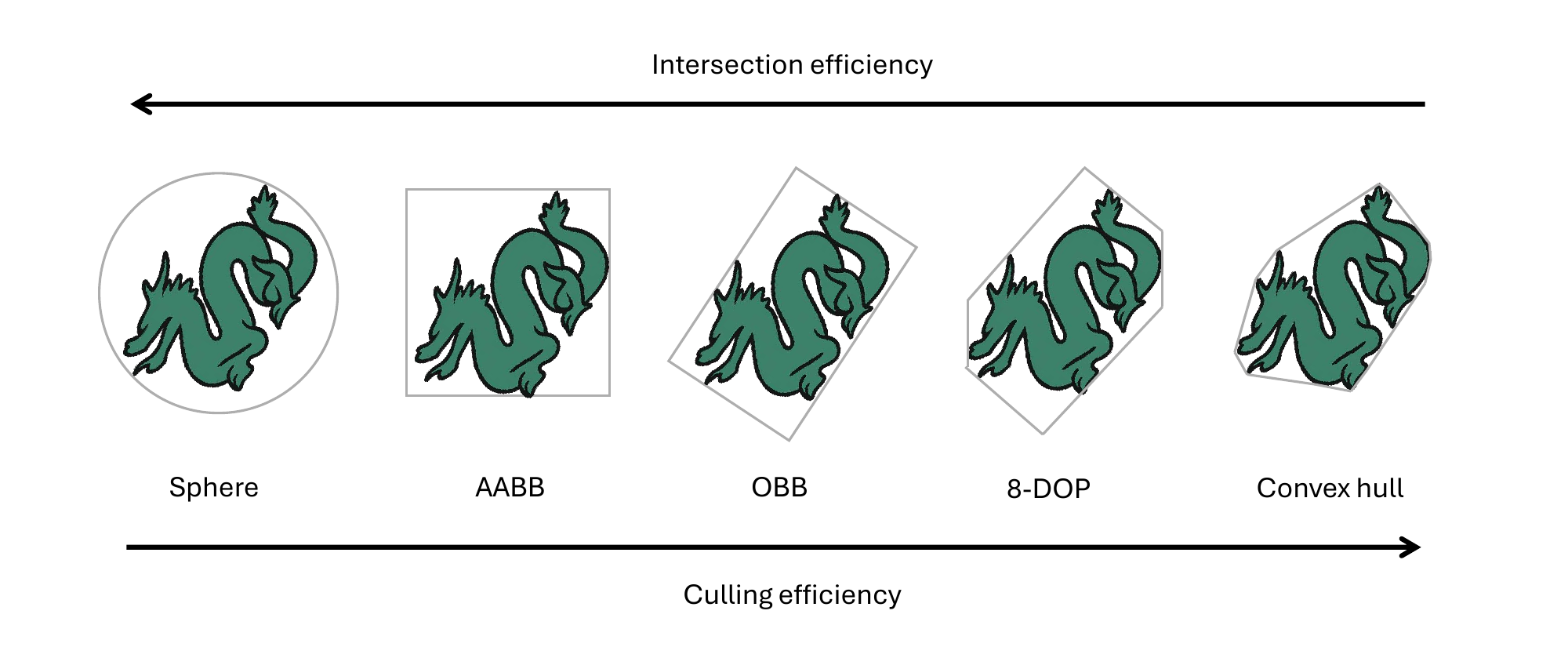}
    \caption{An example of commonly used bounding volumes for ray tracing, balancing trade-off between culling efficiency (tightness) and the intersection efficiency: sphere, axis-aligned bounding box (AABB), oriented bounding box (OBB), discrete oriented polytope (DOP), and convex hull.}
    \label{fig:bvs}
\end{figure}

\subsection{Bounding Volume Hierarchy (BVH)} 
A drawback of bounding volumes is that if we hit a bounding volume, we still need to compute intersections with all objects inside. Bounding volumes can be nested in a hierarchical manner to form a rooted tree with bounding volumes in the internal nodes and triangles in the leaf nodes \cite{meister2021}. We can then find the intersection by traversing the hierarchy from the root, skipping the nodes that are not intersected. While the logarithmic complexity is not guaranteed in general, depending on how the tree is balanced, the complexity is significantly reduced in practice. Contemporary ray tracing frameworks use wide trees with a maximum branching factor of 4 or 8.

\begin{figure}
    \centering
    \includegraphics[width=0.8\linewidth]{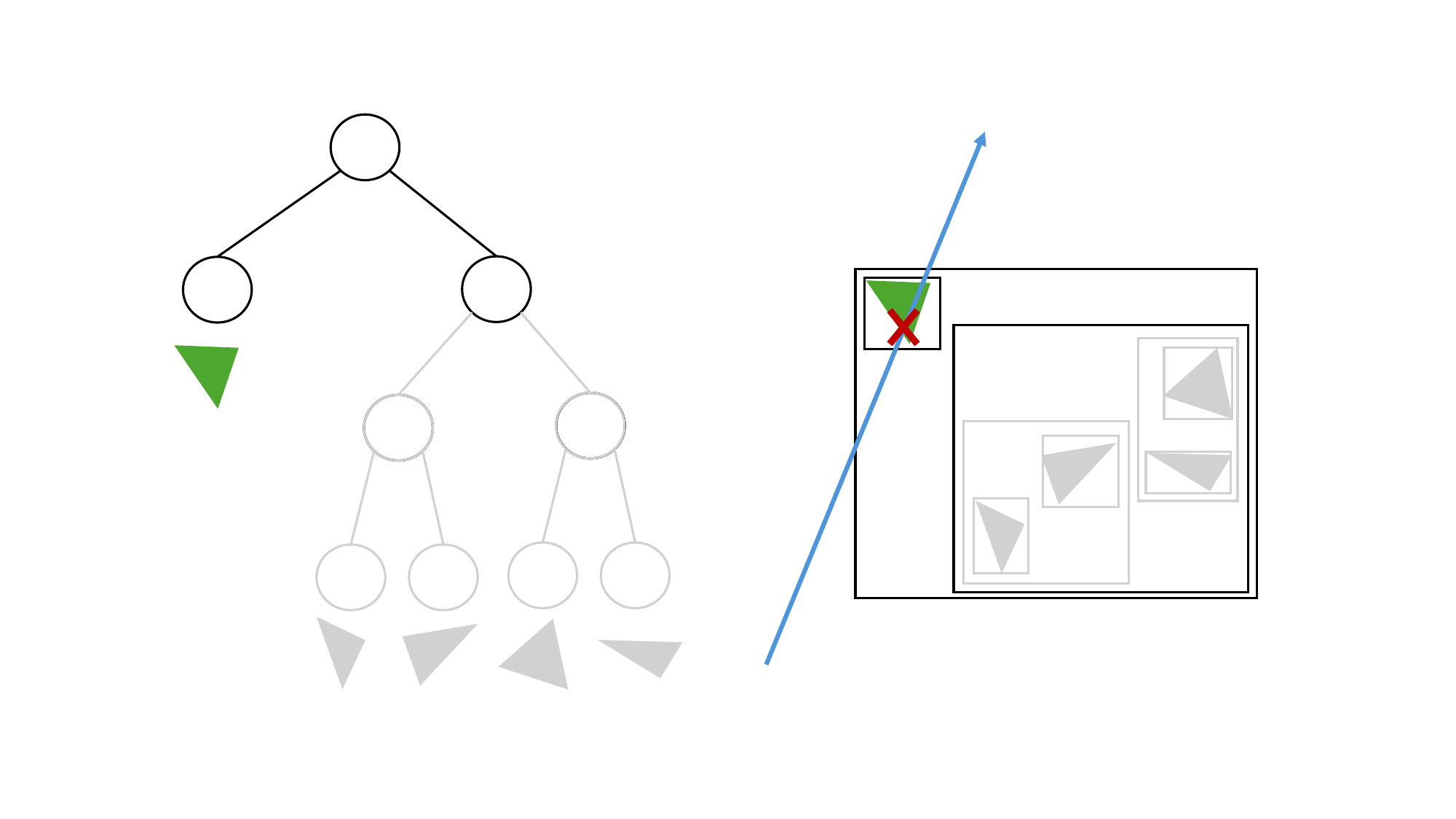}
    \caption{An example a simple bounding volume hierarchy (BVH) using axis-aligned bounding boxes (AABBs). To find the closest intersection (the red cross), we need to test three AABBs and one triangle (the gray part is culled off).}
    \label{fig:bvhs}
\end{figure}

\paragraph{BVH Construction} 
To be able to find the intersection, we need to construct the BVH first. There are two main construction approaches: top-down (splitting)~\cite{wald2007} and bottom-up (clustering)~\cite{walter2008}. The former approach starts with all triangles, and recursively splits them into disjoint subsets until a termination criterion (e.g., the number of triangles in a subset is less than a predefined threshold) is satisfied. The latter approach considers each triangle as a cluster, then iteratively merges the clusters until only one clusters remains, which corresponds to the root node. Notice that there are many possible BVHs for a given scene or model. For example, there are many ways to split the triangles during the top-down construction or to merge the clusters during the bottom-up construction. These local decisions have an impact on culling efficiency, which is inversely proportional to a cost function defined as a sum of surface areas of the bounding volumes in the internal nodes. BVH construction algorithms aim to minimize the surface areas and the construction times, preferring one of these criteria depending on a particular application \cite{meister2021}. The construction speed is critical for dynamic geometry, where once the underlying triangles change, the BVH becomes invalid, and thus must be reconstructed. An alternative approach is to keep the topology of the BVH the same and fit the bounding volumes to the current geometry, which can be done efficiently in a bottom-up manner. Depending on the bounding volume type, we can compute new bounding volumes simply from the child bounding volumes (e.g., for AABBs). A caveat is that the culling efficiency may degrade if the geometry is very different than the original geometry for which the BVH was initially constructed.

\paragraph{BVH Traversal} 
Once the BVH is constructed, it is relatively straightforward to use it to find the intersections. To traverse the BVH, we typically employ a stack to store the nodes to be tested, starting by pushing the root node onto the stack. In each step, we pop a node from the top of the stack and test it, if the bounding volume of the node is intersected, we either push its children onto the stack (in the case of an internal node), or test triangles inside (in the case of a leaf node). We repeat this until the stack is empty. Using this procedure, we can find both the nearest intersection, any intersection, or all intersections.

\paragraph{Two-Level Hierarchy and Instancing}
In practice, it is beneficial to arrange the scene into two levels~\cite{wald2003}, where the bottom level consists of BVHs of individual scene objects (bottom-level acceleration structure - BLAS) and the top-level BVH (top-level acceleration structure - TLAS) is built over these BVHs such that the leaf nodes of the top BVH contain references to the bottom level BVHs and an affine transformation. There are two advantages of this approach. This approach allows instancing of the bottom level objects. We reference the same bottom-level BVH multiple times in the top-level BVH with different transformations (e.g., a classroom with multiple instances of desks and chairs), which reduces the memory usage. Moreover, this approach enables rigid animations by changing the transformations such that only the top-level BVH has to be reconstructed, which significantly reduces the update time in real-time applications. A drawback is that the ray traversal becomes more complicated. When we encounter an instance node during the traversal, we need to transform the ray into the local object space using the inverse transformation.

%% file: sec/7_hiprt.tex
\section{HIPRT: HIP Ray Tracing API}
While implementing a simple variant of BVH is relatively straightforward, optimizing the code for high performance is very challenging. HIPRT~\cite{meister2024} is an open-source ray tracing framework written in HIP, implementing the BVH construction, traversal, and other essential features required in modern renderers. HIPRT is optimized for AMD GPUs (including MI series), utilizing specialized ray tracing hardware units on RDNA 2 and later architectures. In this \articleitself{}, we briefly introduce the HIPRT API, illustrating how to build a two-level hierarchy on a simple example.

\subsection{Context Initialization}
Note that HIPRT performs all computation entirely on the GPU, and thus all input buffers in the following code are supposed to be allocated on the device. We start with the HIPRT context initialization and setting the logging level that may be helpful for debugging (see Listing~\ref{lst:hiprt_context}).

\FloatBarrier
\begin{lstlisting}[caption=Context initialization and log level setting.,style=myHIPStyle2,label={lst:hiprt_context}]
#include <hiprt/hiprt.h>

hiprtContextCreationInput hiprtCtxInput;
hiprtCtxInput.deviceType = hiprtDeviceAMD;
hiprtCtxInput.ctxt = oroGetRawCtx(ctx);
hiprtCtxInput.device = oroGetRawDevice(device);	

hiprtContext hiprtCtx;
hiprtCreateContext(HIPRT_API_VERSION, hiprtCtxInput, hiprtCtx);

hiprtSetLogLevel(hiprtLogLevelError | hiprtLogLevelWarn);
\end{lstlisting}
\FloatBarrier

\subsection{Geometry Construction}
We build a bottom-level BVH for a triangle mesh. We first initialize the HIPRT triangle mesh, geometry build input, and build options structures (see Listing~\ref{lst:geom_build_input}). In the following example, we assume indexed geometry with triangles defined in a consecutive array of \verb|uint3| structures and vertices are stored in a consecutive array of \verb|float3| structures. In the build options, we can specify the construction algorithm and a few other options. In this case, we use the balanced option that provides very good culling efficiency and the build is still very fast (the other two options are fast and high-quality).

\FloatBarrier
\begin{lstlisting}[caption=Triangle mesh and geometry build input.,style=myHIPStyle2,label={lst:geom_build_input}]
hiprtTriangleMeshPrimitive mesh{};
mesh.triangleCount = /* the number of triangles */;
mesh.triangleStride = sizeof(uint3);
mesh.triangleIndices = /* a device pointer to an array of uint3 */;
mesh.vertexCount  = /* the number of vertices */;
mesh.vertexStride = sizeof(float3);
mesh.vertices = /* a device pointer to an array of float3 */;

hiprtGeometryBuildInput geomInput{};
geomInput.type = hiprtPrimitiveTypeTriangleMesh;
geomInput.primitive.triangleMesh = mesh;

hiprtBuildOptions options{};
options.buildFlags = hiprtBuildFlagBitPreferBalancedBuild;
\end{lstlisting}
\FloatBarrier

The construction requires a temporary space for intermediate computations. Based on the mesh size and other options, we query the size of this space and allocate the corresponding temporary buffer on the device (see Listing~\ref{lst:temp_buff}). This buffer can be reused for construction of other BVHs or any other computations on the user side.

\FloatBarrier
\begin{lstlisting}[caption=Temporary buffer allocation.,style=myHIPStyle2,label={lst:temp_buff}]
size_t geomTempSize{};
hiprtGetGeometryBuildTemporaryBufferSize(hiprtCtx, geomInput, options, geomTempSize);

hiprtDevicePtr geomTempTris{};
oroMalloc(reinterpret_cast<oroDeviceptr*>(&geomTempTris), geomTempSize)
\end{lstlisting}
\FloatBarrier

Last, we create and build the HIPRT geometry, which corresponds to the bottom-level BVH in the HIPRT terminology (see Listing~\ref{lst:geom_build}).

\FloatBarrier
\begin{lstlisting}[caption=Geometry creation and construction.,style=myHIPStyle2,label={lst:geom_build}]
hiprtGeometry geomTris{};
hiprtCreateGeometry(hiprtCtx, geomInput, options, geomTris);
hiprtBuildGeometry(hiprtCtx, hiprtBuildOperationBuild,  geomInput, options, geomTempTris, 0 /* stream */, geomTris);
\end{lstlisting}
\FloatBarrier

HIPRT supports custom geometric primitives such as spheres or curves. HIPRT is agnostic to a particular type of the primitive, taking a list of AABBs of these primitives provided by a user as an input (in contrast to the triangle mesh). In the example showed in Listing~\ref{lst:custom_prims}, we assume that each AABB is represented as two \verb|float3| structures. The geometry type is a user defined value that will be later used to set up the custom intersection functions. The rest of the BVH construction remains the same as for triangles above.

\FloatBarrier
\begin{lstlisting}[caption=AABB list and geometry build input.,style=myHIPStyle2,label={lst:custom_prims}]
hiprtAABBListPrimitive list{};
list.aabbCount = /* the number of custom primitives */;
list.aabbStride = 2 * sizeof(float3);
list.aabbs = /* a device pointer to an array of 2 * float3 */;;

hiprtGeometryBuildInput geomInput{};
geomInput.type = hiprtPrimitiveTypeAABBList;
geomInput.primitive.aabbList = list;
geomInput.geomType = 0;

hiprtDevicePtr geomTempCustoms{};
...
hiprtGeometry geomCustoms{};
hiprtBuildGeometry(hiprtCtx, hiprtBuildOperationBuild, geomInput, options, geomTempCustoms, 0 /* stream */, geomCustoms);
\end{lstlisting}
\FloatBarrier

\subsection{Scene Contruction}
With constructed HIPRT geometries, we can build HIPRT scene, which corresponds to the top-level BVH. Similarly to geometries, we need to set up the scene build input, which takes an array of HIPRT instances. In Listing~\ref{lst:instances}, we define two instance objects, and we use them for actual instancing, creating an array of the instance objects, where each object can be repeated arbitrary times.

\FloatBarrier
\begin{lstlisting}[caption=Instance definition.,style=myHIPStyle2,label={lst:instances}]
hiprtInstance instanceTris{};
instanceTris.type = hiprtInstanceTypeGeometry;
instanceTris.geometry = geomTris;

hiprtInstance instanceCustoms{};
instanceCustoms.type = hiprtInstanceTypeGeometry;
instanceCustoms.geometry = geomCustoms;

hiprtInstance instances[] = { instanceTris, instanceTris, ..., instanceCustoms, instanceCustoms };
constexpr size_t INSTANCE_COUNT = sizeof(instances) / sizeof(instances[0]);
\end{lstlisting}
\FloatBarrier

Notice that we set the instance type. HIPRT supports multi-level instancing (i.e., instances of instances). In this \articleitself{}, we limit ourselves to two-level hierarchy, and thus we always use the geometry type. We also need to define transformation for each instance. In HIPRT, the instance transformation can be specified as a transform matrix (\verb|hiprtFrameMatrix|) or by individual transformation components (\verb|hiprtFrameSRT|). Listing~\ref{lst:transformations} illustrates the latter case.

\FloatBarrier
\begin{lstlisting}[caption=Instance tranformations - frames.,style=myHIPStyle2,label={lst:transformations}]
hiprtFrameSRT frames[INSTANCE_COUNT];
for (size_t i = 0; i < INSTANCE_COUNT; ++i)
{
  hiprtFrameSRT& frame = frames[i];
  frame.translation = /* translation component of the i-th frame */;
  frame.scale = /* scale component of the i-th frame */;
  frame.rotation = /* rotation component of the i-th frame */;
}
\end{lstlisting}
\FloatBarrier

With the instances and transformations, we can finally set up the scene build input. In Listing~\ref{lst:scene_build_input}, we allocate the device buffers for the instances and transformation, and copy the data to the device.

\FloatBarrier
\begin{lstlisting}[caption=Scene build input and instance data transfer to a device.,style=myHIPStyle2,label={lst:scene_build_input}]
hiprtSceneBuildInput sceneInput{};
sceneInput.instanceCount = INSTANCE_COUNT;
sceneInput.frameCount = INSTANCE_COUNT;

oroMalloc(reinterpret_cast<oroDeviceptr*>(&sceneInput.instances), INSTANCE_COUNT * sizeof(hiprtInstance));
oroMemcpyHtoD(reinterpret_cast<oroDeviceptr>(sceneInput.instances), instances, INSTANCE_COUNT * sizeof(hiprtInstance));

oroMalloc(reinterpret_cast<oroDeviceptr*>(&sceneInput.instanceFrames), INSTANCE_COUNT * sizeof(hiprtFrameSRT));
oroMemcpyHtoD(reinterpret_cast<oroDeviceptr>(sceneInput.instanceFrames), frames, INSTANCE_COUNT * sizeof(hiprtFrameSRT));
\end{lstlisting}
\FloatBarrier

Once the scene build input is set up, we allocate the temporary buffer for the construction. Finally, we create and build the HIPRT scene (see Listing~\ref{lst:scene_build}).

\FloatBarrier
\begin{lstlisting}[caption=Scene creation and construction,style=myHIPStyle2,label={lst:scene_build}]
size_t sceneTempSize{};
hiprtGetSceneBuildTemporaryBufferSize(hiprtCtx, sceneInput, options, sceneTempSize);

hiprtDevicePtr sceneTemp{};
oroMalloc(reinterpret_cast<oroDeviceptr*>(&sceneTemp), sceneTempSize)

hiprtScene scene{};
hiprtCreateScene(hiprtCtx, sceneInput, options, scene);
hiprtBuildScene(hiprtCtx, hiprtBuildOperationBuild, sceneInput, options, sceneTemp, 0 /* stream */, scene);
\end{lstlisting}
\FloatBarrier

One could wonder why the frames are not stored in the same structure as instances. HIPRT supports instance-based motion blur such that one instance may have a range of frames. The ranges need to be specified as additional buffer in the scene build input (\verb|instanceTransformHeaders|).

\subsection{Custom Intersection}
For the custom primitives, we need to set up the intersection function and also the data defining the custom primitives that are passed to the intersection function. To pass the data, we need to create the function table which is a 2D structure mapping a ray type and geometry type to the corresponding data in the intersection function. An example of this is showed in Listing~\ref{lst:func_table}. We assume a single ray type and geometry type, resulting in a single entry in the table.

\FloatBarrier
\begin{lstlisting}[caption=Custom function table - initialization and data assignment.,style=myHIPStyle2,label={lst:func_table}]
hiprtFuncDataSet funcDataSet{};
funcDataSet.intersectFuncData = /* arbitrary device pointer */;

hiprtFuncTable funcTable{};
checkHiprt(hiprtCreateFuncTable(ctxt, 1 /* the number of geometry types */ , 1 /* the number of ray types */, funcTable));
checkHiprt(hiprtSetFuncTable(ctxt, funcTable, 0 /* geometry type */, 0 /* ray type */, funcDataSet));
\end{lstlisting}
\FloatBarrier

The intersection function must be a device function with the signature showed in Listing~\ref{lst:custom_intersection}, returning \verb|true| in the case of intersection and \verb|false| otherwise. The function takes a ray, the data we specified in the previous step, an optional payload, and the resulting hit structure.

\FloatBarrier
\begin{lstlisting}[caption=Signature of a custom intersection function.,style=myHIPStyle2,label={lst:custom_intersection}]
__device__ bool intersectCustom(const hiprtRay& ray, const void* data, void* payload, hiprtHit& hit) {...}
\end{lstlisting}
\FloatBarrier

\subsection{Traversal Stacks and Traversal Objects}
Listing~\ref{lst:stack} shows how to use the constructed scene to find the intersections. We need to create a traversal object that can be used to find the intersection. HIPRT allows to customize the traversal stack. For high performance, we use the HIPRT global stack that combines local shared memory and global memory as a fallback. The global stack buffer must be big enough to accommodate stacks for all scheduled threads. Similarly, the shared memory buffer must be big enough for all threads in the block.

\FloatBarrier
\begin{lstlisting}[caption=Global traversal stack setup shared and global buffers.,style=myHIPStyle2,label={lst:stack}]
__shared__ uint32_t sharedStackCache[SHARED_STACK_SIZE * BLOCK_SIZE];

hiprtSharedStackBuffer sharedStackBuffer{};
sharedStackBuffer.stackSize = SHARED_STACK_SIZE;
sharedStackBuffer.stackData = sharedStackCache;

hiprtGlobalStackBuffer globalStackBuffer{};
globalStackBuffer.stackSize = /* global buffer size */;
globalStackBuffer.stackData = /* global buffer pointer */;

hiprtGlobalStack stack(globalStackBuffer, sharedStackBuffer);
hiprtEmptyInstanceStack instanceStack;
\end{lstlisting}
\FloatBarrier

There are two main types of traversal objects: for closest intersection and for any/all intersections. Note that all intersections can be found by repeatedly calling \verb|getNextHit()| on the traversal object. We can check the state of the traversal by calling \verb|getCurrentState()| on the traversal object, which indicates whether all intersection have been found. The state of traversal is stored in the stack buffers. For example, if we reuse the buffers in the closest traversal object, we invalidate this state of the any-hit traversal object (as we did in the example above). An example of the traversal objects and using them to find the intersection is depicted in Listing~\ref{lst:traversal}.

\FloatBarrier
\begin{lstlisting}[caption=Using the traversal object to find the intersections.,style=myHIPStyle2,label={lst:traversal}]
hiprtRay ray{};
ray.origin = /* ray origin */ 
ray.direction = /* ray direction */ 
ray.minT = /* min ray length */ MIN_T;
ray.maxT = /* max ray length */ MAX_T;

hiprtSceneTraversalAnyHitCustomStack<hiprtGlobalStack, hiprtEmptyInstanceStack> anyHitTr(scene, ray, stack, instanceStack, hiprtFullRayMask, hiprtTraversalHintDefault, 
    nullptr /* payload */, funcTable);
hiprtHit anyHit = anyHitTr.getNextHit();

hiprtSceneTraversalClosestCustomStack<hiprtGlobalStack, hiprtEmptyInstanceStack> closestTr(scene, ray, stack, instanceStack, hiprtFullRayMask, hiprtTraversalHintDefault, nullptr /* payload */, funcTable);
hiprtHit closestHit = closestTr.getNextHit();
\end{lstlisting}
\FloatBarrier

There are three things in the code we have not explained yet: the instance stack, the ray mask, and the traversal hit. All three are not relevant for our use-case, but they could be useful in more complex scenarios. The instance stack is needed in the case of multi-level instancing. In our case, we use dummy empty stack that disables code for multi-level instancing and let compiler optimize the kernel more efficiently. We can optionally define a mask (32 bits) for each instance in the scene build input. This mask is tested against the ray mask (using the bitwise AND). In our case, as we have not specified these mask, they are assumed to be full such that the test returns always true. Last, the traversal hit can indicate from which distribution the ray has been generated (e.g., a shadow ray). However, in practice, it almost always best to use the default hint.

\subsection{Trace Kernel Compilation}
The last missing piece is the trace kernel compilation, which injects the HIPRT functionality to a user kernel. In principle, the trace kernel is nothing else than a regular HIP kernel. As HIPRT is an open-source project, the obvious way is to include the HIPRT headers directly in the kernel. Another option is to use \emph{bitcode} linking that allows to link pre-compiled HIPRT device code to the user kernel. In general, we can use for the compilation either by \verb|HIPCC| or \verb|HIPRTC| in runtime. 

This approach works fine unless we use custom functions (e.g., the intersection function we defined above). Note that HIPRT supports besides the custom intersection functions also custom intersection filters that allow to filter out intersections during the traversal (e.g., alpha masking or filtering out self-intersections). The intersection filters are user callbacks similar to the custom intersection functions, but they can be used also for triangles compared to the custom intersection. The setup is practically the same as for the custom intersection functions. The custom functions make the compilation more complex as HIPRT needs to generate a function that dispatches custom functions based on the geometry type we set before. If we want to use \verb|HIPCC|, we need to define the dispatch function manually. 

Listing~\ref{lst:tracer_kernel} shows how to compile the trace kernel in runtime using \verb|hiprtBuildTraceKernels| function that builds the dispatch function out of the box. 
First, we create the function name-set and set the name of our custom intersection function. Then, we call \verb|hiprtBuildTraceKernels| to compile the module, generating the dispatch function and injecting the HIPRT traversal code to the user kernel.

\FloatBarrier
\begin{lstlisting}[caption=Trace kernel compilation via the HIPRT API.,style=myHIPStyle2,label={lst:tracer_kernel}]
hiprtFuncNameSet funcNameSet{};
funcNameSet.intersectFuncName = "intersectCustom";

hiprtApiFunction functionOut{};

std::string sourceCode = /* the source code of the HIP module */
std::string functionName = /* kernel function name */

hiprtBuildTraceKernels(
	hiprtCtx,
	1, /* the number of kernel functions */
	functionName.c_str(), /* the names of kernel functions */
	sourceCode.c_str(), /* the source code of the HIP module */
	"", /* path to the module source file (optional) */
	0, /* the number of the included headers (optional) */
	nullptr, /* sources of the included headers (optional) */
	nullptr, /* the names of included header (optional) */
	0, /* the number of compilations options (optional) */
	nullptr, /* compilation options (optional) */
	1, /* the number of geometry types */
	1, /* the number of ray types */
	&funcNameSet, /* custom function namesets (optional) */
	&functionOut, /* resulting HIP functions */
	nullptr, /* resulting HIP module (optional) */
	false /* cache compiled kernes (optional) */);
\end{lstlisting}
\FloatBarrier

The signature of the function is similar to \verb|hiprtcCreateProgram| in \verb|HIPRTC|. There are a few optional arguments providing flexibility of the compilation. The mandatory arguments are as follows: the number of kernels to be compiled, the names of the kernels, and the source code of the module as a string. The compiled functions are in the form of \verb|hiprtApiFunction|, which can be easily cast, for example, to \verb|hipFunction_t| and launched using the standard HIP API.

Custom functions are passed as an array of \verb|funcNameSet| structures. The layout of this array must correspond to the the layout of the function table we defined before. In our example, we have only one entry, and thus we pass just one \verb|funcNameSet| object. For multiple functions, the layout is a flattened 2D table with rows corresponding to ray types and columns to the geometry types stored in row-major order. Additional information about HIPRT can be found on GPUOpen\footnote{\url{https://gpuopen.com/hiprt/}}.

%% file: sec/8_next_steps.tex
\section{Topics Beyond This \articleitself{}}
In this chapter, we presented an introduction to ray tracing with a focus on the GPU implementation in HIP. However, rendering is a vast area, and we touched only a very small fraction. In this section, we refer interested readers to additional topics for further exploration.

\textbf{Rendering Algorithms} \noindent
We presented unidirectional path tracing with the next event estimation that is sufficient for most of the scenarios. Nevertheless, in complex settings, it might be difficult to reach the light source by tracing rays only from camera. For example, if the light source is enclosed by a glass sphere. In such cases, even the next event estimation fails because the light source is not directly visible. To tackle this problem, various bidirectional algorithms have been proposed. Bidirectional path tracing~\cite{veach1997thesis} casts rays from both the camera and the light sources, stochastically connecting these paths. Metropolis light transport~\cite{veach1997thesis} perturbs light paths using Markov chain Monte Carlo. Photon mapping~\cite{jensen2001pm} operates in two phases. In the first phase are cast photons from the light sources and stored in a spatial data structure. During the second phase, the camera rays are cast, collecting photons on the first diffuse surface. Photon mapping is particularly effective for rendering effects such as caustics.

\textbf{Monte Carlo and Sampling} \noindent
An orthogonal approach to improving the efficiency of the light transport simulation is to sample light paths according to a distribution that matches the underlying light transport. This is called importance sampling (IS) in the context of Monte Carlo integration. The goal is to draw samples where the integrand has high values. The challenge is that besides the value we must also compute the corresponding probability density function. We showed a simple technique sampling according to the cosine term in this chapter. Multiple sampling techniques can be combined using multiple importance sampling (MIS)~\cite{veach1997thesis}. There is a way to improve rendering efficiency by changing the sampling points themselves, which is stratified sampling or Quasi-Monte Carlo sequences~\cite{singh19analysis}. Such sequences are spatially distributed across the domain, and they outperform the convergence of Monte Carlo integration over random samples. Another widely-used technique is Russian roulette~\cite{dutre2002gi}, which stochastically terminates paths based on their actual contribution, elegantly avoiding bias compared to a fixed depth while improving overall efficiency. We refer to  \emph{Advanced Global Illumination} by Dutr\'{e} et al.~\cite{dutre2002gi} for more details about these methods. 

\textbf{Ray Tracing} \noindent
Since all these algorithms rely on ray tracing, efficiency can be further improved by optimizing the ray tracing, specifically the BVH. SBVH~\cite{stich2009sbvh}, a BVH construction method using spatial splits, is an approach that allows objects to be split to achieve tighter bounding boxes, albeit at the cost of higher memory consumption and more complex construction. H-PLOC~\cite{benthin2024hploc} is a highly efficient construction algorithm, building BVHs of very good quality. SBVH and H-PLOC are implemented as high-quality and balanced options, respectively, in HIPRT.

\textbf{Increasing Realism} \noindent
What is important alongside efficiency is the degree of realism. Material properties of surfaces have significant impact on realism, and thus modeling the appearance of real-world materials~\cite{guarnera2016brdf} is another active area of research. In a nutshell, the models describe how the light is reflected (or refracted) from different directions. Volumetric rendering~\cite{novak2018volumes} accounts for optical effects caused by, for instance, fog or smoke, bringing realism to another level. Spectral rendering~\cite{weidlich2021course} models wavelength-dependent effects such as dispersion and diffraction. We refer to \emph{Physically-based Rendering}  by Pharr et al.~\cite{pharr2023} that presents comprehensive overview of the techniques discussed so far. 


\textbf{Real-time Rendering} \noindent
Ray tracing was originally designed for offline rendering, but thanks to advances in both hardware and software, ray tracing is gradually penetrating to real-time applications such as computer games. In this last paragraph, we discuss specifics for real-time scenarios. The major challenge in real-time ray tracing is that time budget is very limited to maintain real-time frame rates, allowing to trace only very few rays in each frame. Therefore, we need to use different tricks to get plausible results in several miliseconds. Denoising and upscaling~\cite{kazmierczyk2025nssd} help mitigate noise caused by insufficient samples and enhance resolution, both of which nowadays rely predominantly on deep learning trained offline on extensive datasets. Radiance caching~\cite{krivanek2005rc} stores radiance in a spatial data structure (or a neural network~\cite{muller2021real}), allowing to terminate light path earlier by approximating the remaining radiance by the cached values. ReSTIR~\cite{bitterli2020restir, wyman2023course} is an algorithm for sampling a large number of light sources designed for real-time scenarios, resampling light candidates from neighboring pixels and form previous frames, exploiting spatiotemporal coherence. 




